\documentstyle[epsf,a4]{article}
\begin{document}
\title{Elliptic flow of $\eta$ and $\pi^{0}$ Mesons \\
in Heavy-Ion Collisions at 2 AGeV }

\author{A. Taranenko$^1$, A. Kugler$^1$, R. Pleska\v{c}$^1$, P. Tlust\'y$^1$,
V. Wagner$^1$, \\
H. L\"ohner$^2$, R. W. Ostendorf$^2$, R.H. Siemssen$^2$, P. H. Vogt$^2$, H.W. Wilschut$^2$, \\
R. Averbeck$^3$, S. Hlav\'{a}\v{c}$^{3,\dagger}$, 
R. Holzmann$^3$,  A.Schubert$^3$, R. S. Simon$^3$, \\
R. Stratmann$^3$, F. Wissmann$^3$, Y. Charbonnier$^4$, 
G. Mart\'{\i}nez$^{4,5}$, Y. Schutz$^4$, \\
J. D\'{\i}az$^5$, A. Mar\'{\i}n$^{5}$, A. D\"{o}ppenschmidt$^6$,
M. Appenheimer$^7$, V. Hejny$^7$, \\ 
V. Metag$^{7}$, 
R. Novotny$^7$, H. Str{\"o}her$^7$, 
J. Wei{\ss}$^7$, A.R. Wolf$^7$ and  M. Wolf$^7$ \\
\\
$^1$Nuclear Physics Institute, CZ-25068 \v{R}e\v{z}, Czech Republic\\
$^2$Kernfysisch Versneller Instituut, NL-9747 AA Groningen, The Netherlands\\
$^3$Gesellschaft f{\"u}r Schwerionenforschung, D-64291 Darmstadt, Germany\\
$^4$Grand Acc\'{e}l\'{e}rateur National d'Ions Lourds, F-14021 Caen
Cedex, France\\
$^5$Instituto de F\'{\i}sica Corpuscular, 
Centro Mixto Universidad de Valencia--CSIC, \\ E-46100 Burjassot, Spain\\
$^6$Institut f{\"u}r Kernphysik, Universit{\"a}t Frankfurt, \\ D-60486
Frankfurt am Main, Germany\\
$^7$II. Physikalisches Institut, Universit{\"a}t Gie{\ss}en,
D-35392 Gie{\ss}en, Germany\\
\\
presented at V TAPS workshop, \v{R}e\v{z}, September 4-8. 1999 \\
}

\normalsize

\date{Report \'UJF-EXP-99/2}
%\date{\today} 

\maketitle

\begin{abstract}
Azimuthal distributions of $\eta$ and $\pi^{0}$ mesons emitted at midrapidity 
in
collisions of 1.9 AGeV $^{58}$Ni+$^{58}$Ni and  2 AGeV $^{40}$Ca+$^{nat}$Ca are
studied as a function of the number of projectile-like spectator nucleons. The
observed anisotropy corresponds to a negative elliptic flow signal for 
$\eta$ mesons, indicating a
preferred emission perpendicular to the reaction plane. 
In contrast, only small azimuthal anisotropies are observed 
for $\pi^{0}$ mesons.
This may indicate that $\eta$ mesons freeze out earlier from the fire ball
than pions.

\end{abstract}

\section{Introduction}
Heavy-ion collisions at incident energies of 1-2 AGeV (Bevalac/SIS energy regime) 
are a unique tool to study nuclear
matter at high density and temperature. According to various theoretical 
model calculations \cite{aich91,cas90,bass98} nuclear matter can be compressed to 2-3 times the
normal nuclear density and heated to temperatures in the order of 100 MeV in this energy
regime. In addition, a fraction of 10-30$\%$ of the participating nucleons is excited to 
short-lived resonance states, mainly $\Delta(1232)$ and $N^*(1535)$ resonances, 
 which subsequently decay via meson emission \cite{ehe93,metag}. Thus the meson observables
 can provide information on the dynamical evolution of the resonance population
 in compressed and excited nuclear matter. 

 The production and decay of the $\Delta(1232)$ resonance is responsible for 
 the production of pions, which are the most 
 abundantly produced secondary particles in the 1-2 AGeV energy regime \cite{har87,bass95,teis97}. 
 A clean signature for the excitation of a high 
 lying resonance is the detection of $\eta$ mesons \cite{metag}. They originate almost exclusively 
 from the decay 
 of the $N^*(1535)$ resonance and thus are sensitive to the abundance of this resonance \cite{br2,gwolf}.
 After the initial production through resonance decays,  pions and $\eta$ mesons
 strongly interact with the surrounding hot nuclear matter in the interaction region and in 
 cold spectator matter. The interactions in hot nuclear matter are the result of 
a complex cyclic process of generation, absorption, and re-emission of mesons 
\cite{bass98,gwolf}, while in cold nuclear matter absorption dominates. 
These processes will influence experimental observables like
the meson abundance as well as the azimuthal angle distributions of mesons
with respect to the reaction plane. 

Azimuthal anisotropy in the emission
of pions in symmetric as well as in asymmetric non-central heavy-ion collisions is a 
clearly established effect and was observed in a wide range of beam energies 
from 0.1 A GeV to 160 A GeV 
%\cite{gos89,kugp94,schu94,bad97,lars,brill,brill97,kint,bar97,app98}.
\cite{gos89} -- \cite{app98}.
Similar to the anisotropy of baryons \cite{rei97}, the effect is usually  
discussed in terms of directed and elliptic flow \cite{olit,vol96}. In a Fourier 
expansion of the azimuthal angle  distribution of particles 
$N(\Delta\varphi$) with respect to the reaction plane 
\begin{equation}
\label{fourier}
N(\Delta\varphi)=v_{0}\left( 1+2\sum_{n\ge 1} v_n\cos n\Delta\varphi\right)
\end{equation}
the first term ($\sim\langle cos\Delta\varphi\rangle$) corresponds to the directed flow, 
while the second term ($\sim\langle cos2\Delta\varphi\rangle$) represents the elliptic 
flow \cite{vol96}. At midrapidity the directed flow $\langle
cos\Delta\varphi\rangle$ of emitted particles vanishes for symmetry reasons and 
only the elliptic flow is present \cite{olit}. 

 In the Bevalac/SIS energy regime the elliptic flow of participant baryons 
 emerging from the collisions at midrapidity was found to be oriented 
perpendicular to the reaction plane (negative elliptic flow,  
$\langle cos2\Delta\varphi\rangle<0$) \cite{kug94, gut90,lei93,brill96}. This effect was
interpreted as a dynamical squeeze-out of nuclear matter due to the build-up 
of pressure in the interaction zone between two colliding nuclei \cite{sto82}.  
Similar to baryons, negative elliptic flow was observed for high transverse 
momentum neutral and charged pions emitted at midrapidity in 1 AGeV Au+Au 
collisions at SIS (GSI) \cite{lars,brill}.
However, the observed pion anisotropy was not been attributed to the expansion of nuclear 
matter, but rather to the strong final state interactions  of 
pions with  cold spectator matter transiently concentrated in the reaction plane
("shadowing"). Consequently, the chance for pion absorption 
is higher in the reaction plane than out-of-plane 
\cite{bass93,bao94}. 
The negative elliptic flow of high energy pions observed in the Bevalac/SIS 
energy regime indicates that they freeze-out while
the spectators are still close to the participant zone \cite{sen99}.  At 
ultrarelativistic  energies the time required for spectator fragments to pass 
the reaction zone (passage time) is so short that the shadowing effect is 
reduced. In this case the geometry of the participant zone favors the
preferential in-plane emission of pions \cite{olit,liu99,bao99} and hence 
positive elliptic flow, which
was observed recently at AGS \cite{bar99} and SPS \cite{agg99,app98} energies. 
However already at SIS  in-plane emission of low energy pions were observed, 
see \cite{lars}. Hence similar scenario can hold for low energy pions even
at SIS energies.

\begin{figure}[h]
 \vspace{-1.0cm}
 \begin{center}
    \mbox{
     \epsfxsize=11.6cm
     \epsffile{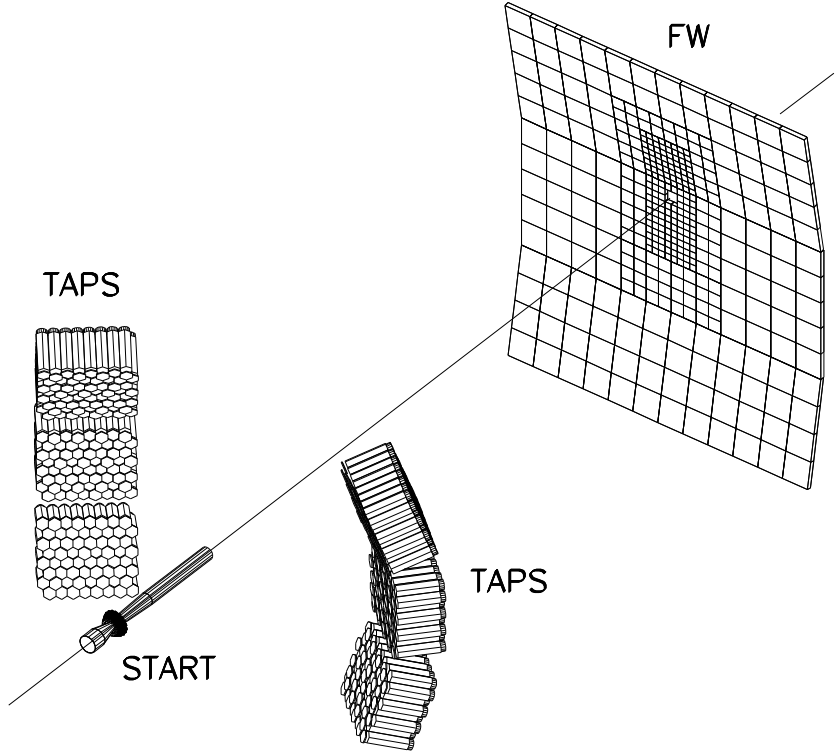}
      }
 \end{center}
 \vspace{-1.0cm}
 \begin{center}\parbox{13cm}
 {
 \caption{ Schematic view of the experimental setup with TAPS and the
KaoS Forward Wall.}
 \label{fg:setup}
 }
 \end{center}
 \vspace{0.0cm}
\end{figure}

\begin{figure}[h]
  \vspace{-1.2cm}
  \begin{center}
    \mbox{
     \epsfxsize=11.6cm
     \epsffile{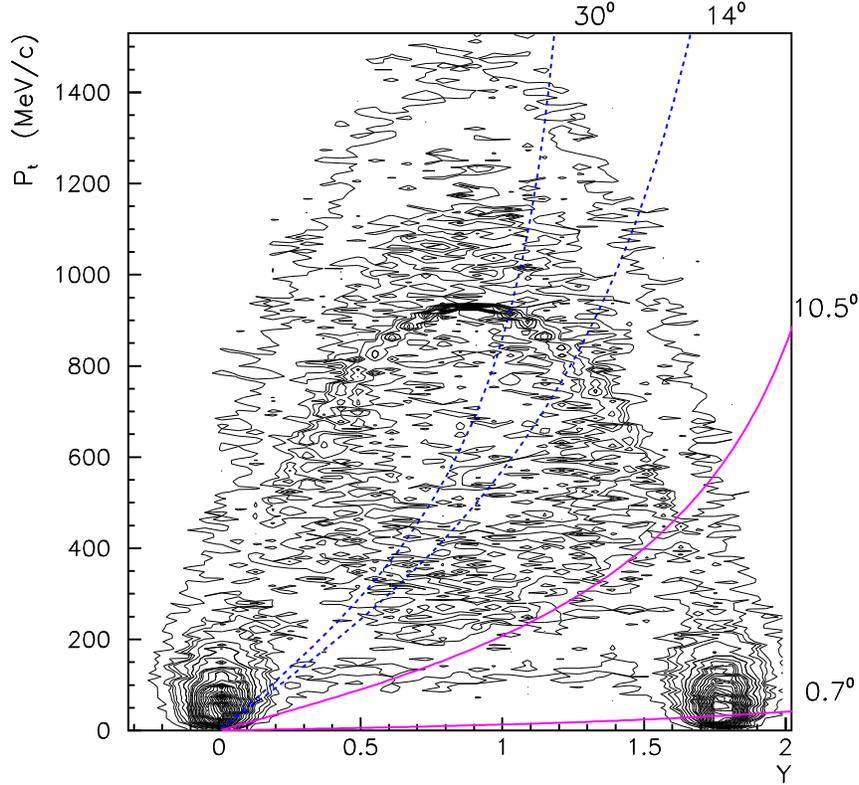}
      }
  \end{center}
  \vspace{-1.0cm}
  \begin{center}\parbox{13.7cm}{
  \caption{The phase space ( transverse momentum p$_t$ versus rapidity Y) for baryons from 
the $^{40}$Ca+$^{40}$Ca at 2.0 A GeV collisions calculated with FREESCO model \cite{fai86}.
The reaction detector covered the polar angles $\Theta_{Lab}$ from 14$^{\circ}$ to 30$^{\circ}$ (dotted lines).
The solid lines indicate the region of the acceptance of the FW detector  
($\Theta_{Lab}$=0.7-10.5$^{\circ}$).
 The rapidity is the laboratory rapidity.
The center of mass rapidity is y$_{cm}$=0.882.}
  \label{fg:freesco}
  }\end{center}
  \vspace{-0.5cm}
\end{figure}
 
\begin{figure}
  \vspace{-.9cm}
  \begin{center}
    \mbox{
     \epsfxsize=9.3cm
     \epsffile{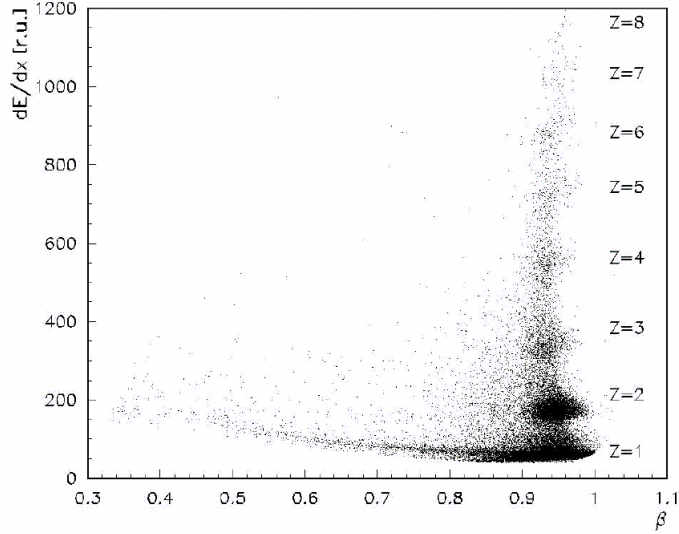}
      }
  \end{center}
  \vspace{-0.6cm}
  \begin{center}\parbox{14.0cm}{
    \caption{The dependence of energy loss dE/dx as a function of particle velocity $\beta$ 
is shown for fragments detected in a single FW module. The data are from the $^{58}$Ni+$^{58}$Ni
reaction at 1.9 AGeV.}
    \label{fg:dE_dx}
   }\end{center}
   \vspace{0.0cm}
\end{figure}

\begin{figure}
  \vspace{-1.3cm}
  \begin{center}
    \mbox{
     \epsfxsize=10cm
     \epsffile{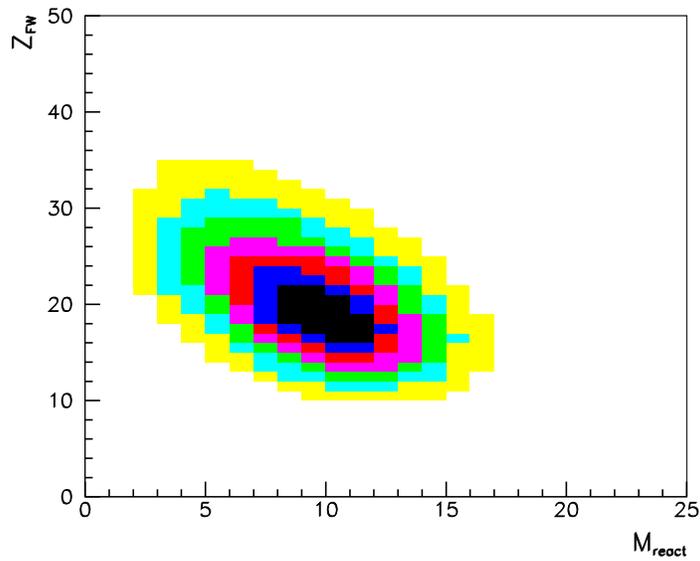}
      }
  \end{center}
  \vspace{-1.1cm}
  \begin{center}\parbox{14.6cm}{
     \caption{The total charge Z$_{FW}$ of particles detected in the FW as a function of the 
charged-particle multiplicity M$_{react}$, measured in coincidence with two neutral hits 
with an energy above 90 MeV in two different TAPS blocks. The data are from the 
$^{58}$Ni+$^{58}$Ni reaction at 1.9 AGeV.}
     \label{fg:sumch2}
   }\end{center}
   \vspace{-0.5cm}
\end{figure}

\begin{figure}
  \vspace{-1.8cm}
  \begin{center}
    \mbox{
     \epsfxsize=9.7cm
     \epsffile{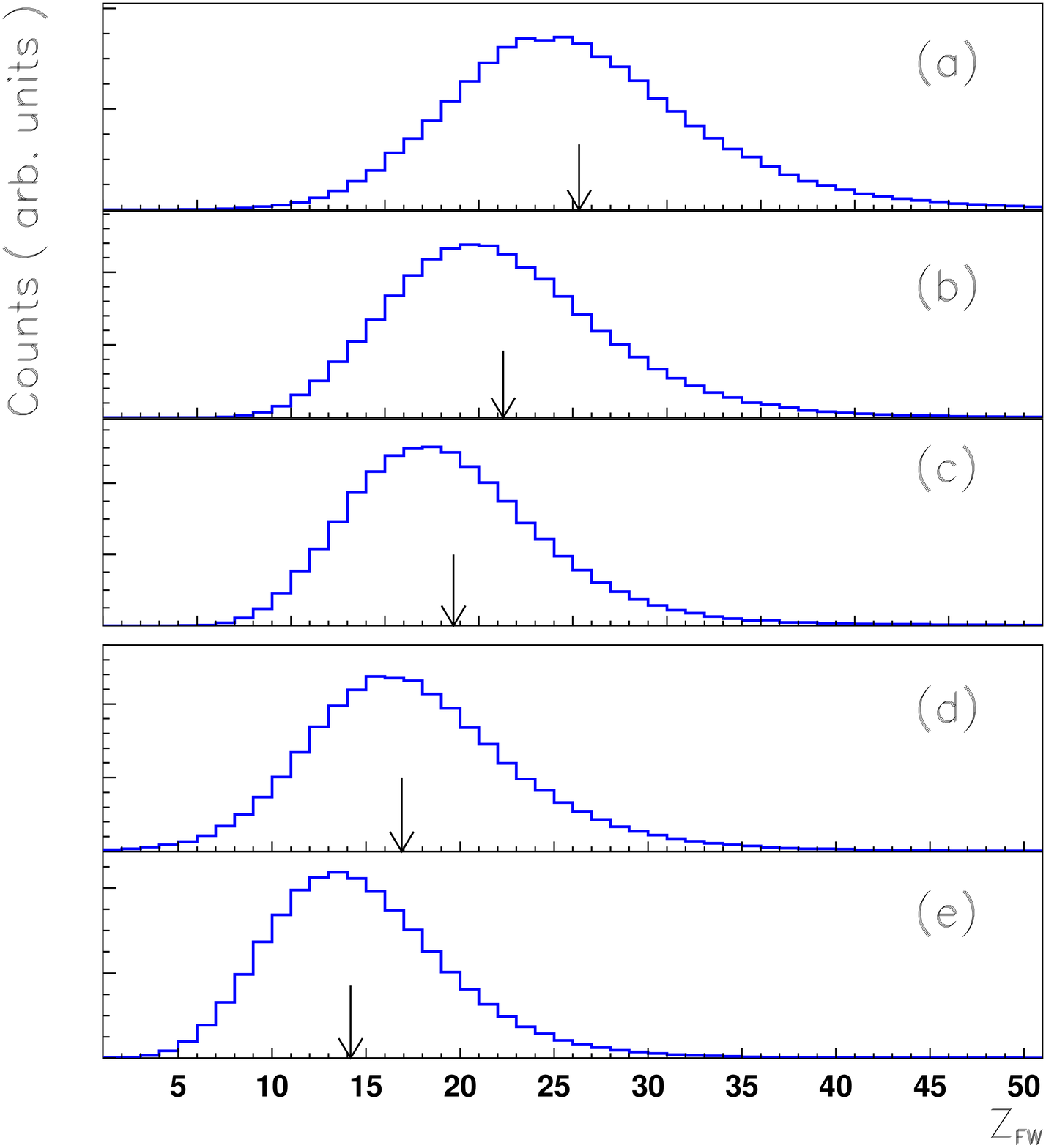}
      }
  \end{center}
  \vspace{-1.1cm}
  \begin{center}\parbox{14.6cm}{
     \caption{The distribution of the total charge of particles detected in the FW for different 
bins in charged-particle multiplicity M$_{react}$ giving the reaction centrality.
(a) - (c) for the experiment $^{58}$Ni+$^{58}$Ni at 1.9 AGeV; (d) - (e) for the
experiment $^{40}$Ca+$^{40}$Ca at 2 AGeV. The arrows indicate the mean values
of the total charge.}
     \label{fg:sumchar}
   }\end{center}
   \vspace{-0.7cm}
\end{figure} 

The elliptic flow of pions is well established and can qualitatively be
described in the frame of transport model calculations  
\cite{bass93,bao94,liu99,bao99}, but the elliptic flow of $\eta$ mesons has not been
observed before. While the absorption of pions in hot nuclear matter proceeds 
via the $\Delta$(1232) resonance which decays dominantly by pion emission, only 
about 50$\%$ of all N$^*$(1535) resonances excited by $\eta$-meson
absorption will reemit $\eta$ mesons \cite{gwolf}. However, the mean free path 
of both mesons in cold nuclear matter is comparable \cite{mami96,paoli89}. 
Therefore, a comparison of the $\eta$- and $\pi^{0}$-elliptic flow may yield 
information on the propagation of these  mesons, as well as on the 
dynamics of the parent baryon resonances.

Moreover, $\eta$ and $\pi^{0}$ mesons can be measured
simultaneously in the same experiment via their two-photon decay channel
(branching ratios: 39.3$\%$ and 98.8$\%$ for $\eta$ and $\pi^{0}$, respectively
\cite{br2}).
Below we present results of the first experimental study of azimuthal 
distributions of $\eta$ mesons emitted in collisions of 1.9 AGeV 
$^{58}$Ni+$^{58}$Ni and 2 AGeV  $^{40}$Ca+$^{nat}$Ca nuclei
and compare them with azimuthal distributions of $\pi^0$ mesons
in the same colliding systems.

\section{ Experiment}

The experiments were performed at the Heavy-Ion Synchrotron SIS at GSI 
Darmstadt. The experimental setup  is sketched in Fig.~\ref{fg:setup}.
In the first experiment a 1.9 AGeV $^{58}$Ni beam with an intensity  
of $6.5\times10^{6}$ particles per spill (spill duration 8~s and repetition 
rate 15~s) was incident on a $^{58}$Ni target (502 mg/cm$^{2}$). In the second
experiment, a $^{nat}$Ca target (320 mg/cm$^{2}$) was bombarded  by a 
$^{40}$Ca beam with kinetic energy 2 AGeV and an intensity of $5\times10^{6}$ particles per
spill (spill duration 10~s and repetition rate 14~s).

Photon pairs from the neutral-meson decay were detected in 
the Two-Arm Photon Spectrometer (TAPS) \cite{taps}. This detector system consisted of 
384 BaF$_{2}$ scintillators arranged in 6 blocks of 64 modules with individual 
Charged Particle Veto detectors (CPV) in front of each module. The blocks were mounted in 
two towers positioned at 40$^{\circ}$ with respect to the 
beam direction at the distance of 150~cm. Three blocks were positioned in each 
tower at +21$^{\circ}$, 0$^{\circ}$ and -21$^{\circ}$ with respect 
to the horizontal plane. In this setup, only neutral mesons around mid-rapidity $y_{cm}$ were
detected. The geometrical acceptance of TAPS 
for the $\pi^{0}$ and $\eta$ detection was roughly $1\times10^{-3}$.

An in-beam plastic scintillator (BC-418) of 200 $\mu$m thickness was used to provide a 
time-zero signal for the time-of-flight measurements as well as to count beam
particles. The plastic Forward Wall (FW) of the KaoS collaboration, see \cite{kaos},
comprising 380 plastic scintillators (BC408)  was positioned 520 cm downstream of the target.
The modules  have a thickness of 2.54 cm with sizes 4x4 cm$^2$ in the center, followed 
by 8x8 cm$^2$ and 16x16 cm$^2$ elements in the outer region. Four modules in the 
center of the wall are removed for a beam pipe made of a carbon fibre tube (diameter 70 mm,
thickness 1 mm). The actual measurements were performed with 320
detectors to obtain a nearly azimuthally symmetric coverage. 
In this arrangement the FW covered the polar angles from 0.7$^{0}$ to 10.5$^{0}$.  
Particles emitted in this angular range are predominantly 
projectile-like spectator nucleons, see Fig.~\ref{fg:freesco}. 
The FW provided the information on emission angle, multiplicity, charge  and time-of-flight 
of protons and light charged fragments up to Z=8, see Fig.~\ref{fg:dE_dx}. 

The reaction detector, comprising 
40 small plastic scintillators (NE102A), was positioned close
to the target and covered the polar angles from 14$^{\circ}$ to 30$^{\circ}$. 
Most of the 
particles emitted in this angular range are  participant nucleons, see
Fig.~\ref{fg:freesco}.
For the selection of events according to the reaction centrality and for the estimation of the
average number of projectile-like spectator nucleons the information
from the reaction detector and the FW were used. The performance of these two detectors is shown
in Fig.~\ref{fg:sumch2}, where the total charge Z$_{FW}$ of the particles detected by the FW is plotted 
as a function of the charged-particle multiplicity M$_{react}$ in the reaction detector, 
measured in coincidence with two neutral hits in different blocks of TAPS.
 
Events were selected according to the reaction centrality by requiring equal statistics of 
mesons in each bin, ranging from 
peripheral reactions with low multiplicity M$_{react}$ to central reactions with high 
multiplicity M$_{react}$.
The total charge Z$_{FW}$ of particles detected by the FW allowed us to estimate the 
mean number of projectile-like spectators $\langle A_{sp}\rangle $ for each studied 
bin in M$_{react}$ as determined by the reaction detector. We used  the relation 
\begin{equation} 
\langle A_{sp} \rangle=\langle Z_{FW}\rangle A_{proj}/Z_{proj},
\end{equation} 
where $A_{proj}$ and $Z_{proj}$ are the mass number and charge of the projectile, 
respectively. The distributions of the total charge $Z_{FW}$ are shown
in Fig.~\ref{fg:sumchar} for both the Ni+Ni and Ca+Ca collisions. The resulting values 
of the mean number of projectile-like spectators $\langle A_{sp}\rangle $ are listed in
the Table~\ref{tb:Asp}. The systematic error of the values $\langle A_{sp} \rangle$ was found \cite{wolf98}
to be less than 4 units. 

\begin{table}
\begin{center}
\begin{tabular}{|c|c|c|c|c|c|} \hline
Reaction &\multicolumn{3}{|c|}{$^{58}$Ni+$^{58}$Ni at 1.9 A GeV}
&\multicolumn{2}{|c|}{$^{40}$Ca+$^{nat}$Ca at 2$A$ GeV}\\ \hline
 M$_{react}$ & 2 - 6 & 7 - 10 & $\ge$ 11 & 1 - 3 & $\ge$ 4\\ \hline
$\langle A_{sp} \rangle$    & 51  & 44 & 37 &  34 & 28\\ \hline
\end{tabular}
  \caption{ The mean number of projectile-like spectators $\langle A_{sp} \rangle$  
for studied bins in the charged-particle multiplicity M$_{react}$ measured by the reaction 
detector.}
  \label{tb:Asp}
\end{center}
\end{table} 

Neutral-meson candidate events were selected by special triggers exploiting
the kinematical constraints of meson decay into photons. For each TAPS block the
hit information for the BaF$_2$ signals above a certain threshold energy and for
the corresponding CPV modules were fed to a multiplicity/pattern unit. This unit performed an 
on-line selection of events induced either by a charged particle (if the BaF$_2$ scintillator 
and the corresponding CPV module have fired simultaneously) or by a neutral particle 
(if the BaF$_2$ scintillator has fired without a corresponding CPV signal).
For the analysis of the present data the following triggers were chosen:\\
For the reaction  $^{40}$Ca+$^{nat}$Ca at 2.0 A GeV the event-selection trigger required \\
a) hits with an energy E$\ge$15 MeV in any two TAPS blocks for the $\pi^{0}$ measurement, and\\
b) at least one neutral hit in each tower with an energy E$\ge$90 MeV for $\eta$ mesons.\\
For the reaction  $^{58}$Ni+$^{58}$Ni at 1.9 A GeV the trigger required \\
a) one neutral hit with an energy E$\ge$90 MeV and one hit with an energy E$\ge$15 MeV in any 
two TAPS blocks for a $\pi^{0}$ candidate, and\\
b) two neutral hits with an energy E$\ge$90 MeV in any two TAPS blocks for selection of 
$\eta$ meson candidate events.\\
In addition, all triggers mentioned above required the coincidence with a signal from the 
reaction detector and the FW. \\
In both reactions downscaled data were taken corresponding to minimum bias trigger requiring
coincidence between the reaction detector and the FW.

\section{ Data Analysis and Simulation}
\subsection{Reaction-plane determination}

The study of event anisotropy requires the determination of the orientation of the reaction plane.
With respect to this plane the preferred emission of some subset of particles can be analysed.
The reaction plane was determined by a modified version \cite{fai87} of the transverse-momentum
method \cite{trmom}. For each event the reaction plane was defined by the incident beam
direction and the vector $\vec{Q}$, which is the weighted sum of the transverse momentum vectors 
of all charged particles detected in an event. Since almost all particles detected by the FW 
have nearly the same velocity (see Fig.~\ref{fg:dE_dx}.), the transverse velocity vector $\vec{v}_{k}^{\perp}$
will be oriented parallel to the position vector $\vec{r}_{k}$ of the particle $k$ in the 
x-y plane (perpendicular to the beam direction) of the FW. 
Here it is assumed that the beam position in the plane of the FW is (0,0). 
Therefore, in our analysis the  vector $\vec{Q}$ is defined as
\begin{equation}
\vec{Q}=\sum^{M}_{k=1}\omega_{k}
\frac{\vec{r}_{k}}{\mid \vec{r}_{k}\mid} ,
\end{equation}
where the sum runs over all $M$ particles detected by the FW in the event, 
$\vec{r}_{k}$ is the position vector of particle $k$ in the x-y plane and 
$\omega_k$ is the weight factor, which depends
on the rapidity of the emitted particle. Since the coverage of the FW excludes nearly all 
particles with $y_{k}<y_{cm}$ the position vectors in Eq. 3 will be summed with a positive 
weight $\omega_{k}$. In order to provide the best reaction plane resolution the weight factor 
$\omega_{k}$ was chosen to be $\omega_{k}$=$Z_{k}$, where $Z_k$ is the charge
 of particle $k$ detected by the FW. 
 
\begin{figure}
  \vspace{-1.2cm}
  \begin{center}
    \mbox{
     \epsfxsize=13.5cm
     \epsffile{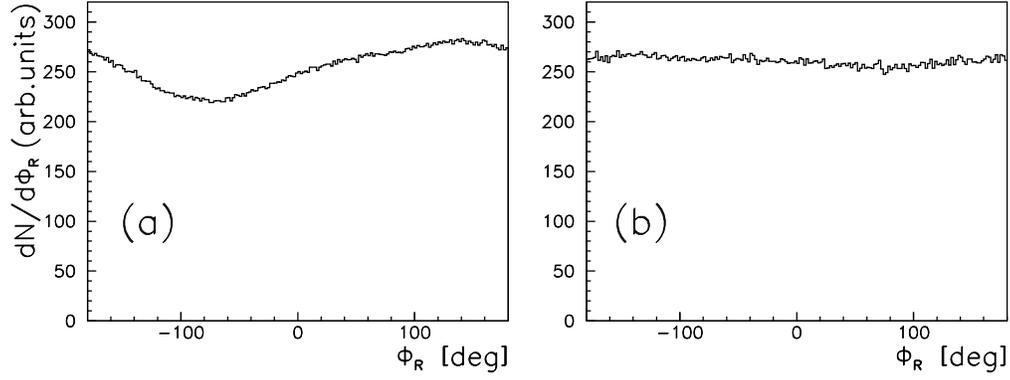}
      }
  \end{center}
  \vspace{-0.7cm}
  \begin{center}\parbox{14.0cm}{
     \caption{
The distributions  of the reaction plane angle $\Phi_{R}$ for peripheral ( 2$\le M_{react} \le $6 )
$^{58}$Ni+$^{58}$Ni reactions before (a) and after (b) the beam-offset correction.}
     \label{fg:offset}
  }\end{center}
  \vspace{-0.8cm}
\end{figure}

Since the coverage of the FW does not overlap
with the TAPS spectrometer (used for the neutral meson reconstruction), the analysis of
azimuthal angle distributions of $\eta$ and $\pi^{0}$ mesons with respect to the reaction
plane is free from autocorrelation effects.
 The analysis of azimuthal anisotropies is rather sensitive to different kinds of experimental
biases, which could simulate an event anisotropy. Such artificially created anisotropies can be 
removed by requiring the distribution of the reaction plane to be isotropic for unbiased events.
The off-line data analysis revealed that the actual beam position in the plane of the FW 
with respect  to the geometrical center of the FW varied with time. Consequently, the
distribution of the reconstructed reaction-plane angle is not flat. This effect required a
correction of the vector 
$\vec{r}^{corr}_{k}=\vec{r}_{k} - \vec{r}_{off}$ for each particle in an event. 
The  offset values
($\langle x_{off} \rangle$, $\langle y_{off} \rangle$) were determined by averaging
over event samples in fixed time intervals. The required corrections 
($\langle x_{off} \rangle$, $\langle y_{off} \rangle$) stayed below 1 cm. As an example of 
this procedure, Fig.~\ref{fg:offset} shows the distributions of the reaction-plane angle ($\Phi_{R}$) 
for peripheral ($2\le M_{react}\le 6$) Ni+Ni reactions at 1.9 AGeV before and 
after including the beam-offset correction. The corrected distribution
(Fig.~\ref{fg:offset}b) is free
from significant distortions which could influence the study of azimuthal anisotropies.

\subsection{Reaction-plane resolution}

\begin{figure}
  \vspace{-1.0cm}
  \begin{center}
    \mbox{
     \epsfxsize=9.0cm
     \epsffile{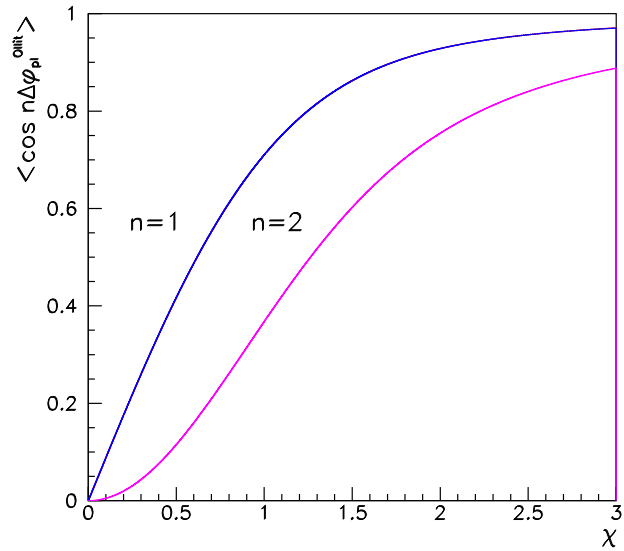}
      }
  \end{center}
  \vspace{-1.0cm}
  \begin{center}\parbox{13.8cm}{
     \caption{The variation of $\langle cos~n\Delta\phi_{pl}^{Ollit}\rangle$
with the parameter $\chi$, calculated from Eq. 5 for the first two harmonics n=1 and n=2.}
     \label{fg:bess}
  }\end{center}
  \vspace{0.0cm}
\end{figure}  

\begin{figure}
  \vspace{-1.2cm}
  \begin{center}
    \mbox{
     \epsfxsize=14.0cm
     \epsffile{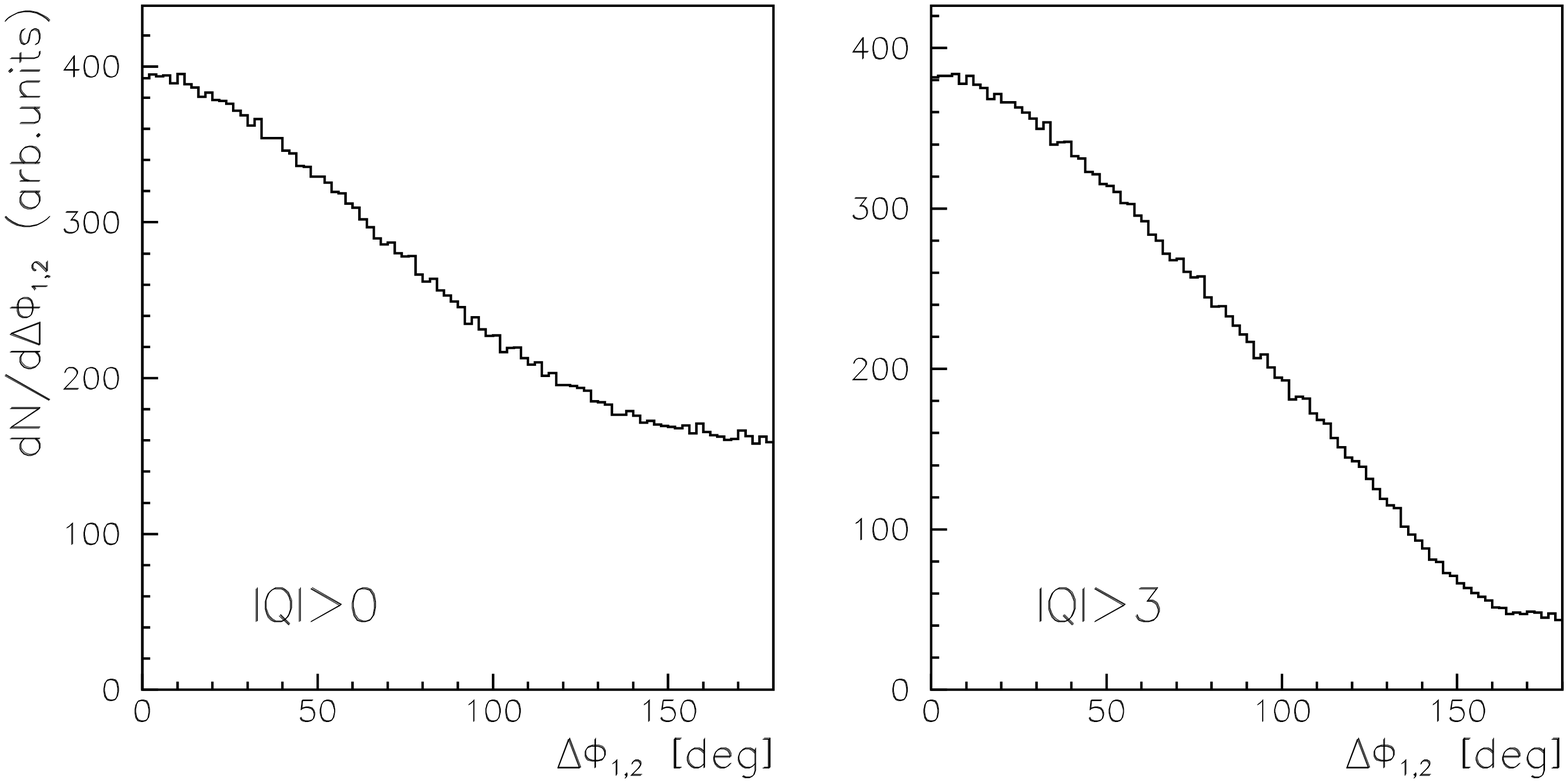}
      }
  \end{center}
  \vspace{-0.5cm}
  \begin{center}\parbox{13.6cm}{
     \caption{
The distribution of the relative azimuthal angle $\Delta \Phi_{1,2}$ between the reaction planes 
$\vec{Q_1}$ and $\vec{Q_2}$ of two subevents. The right part of the figure shows data for a well defined reaction
plane with $\mid\vec{Q}\mid > 3$. The data are from peripheral (2$\le M_{react} \le $6)
$^{58}$Ni+$^{58}$Ni reactions.}
     \label{fg:rplane}
  }\end{center}
  \vspace{-0.0cm}
\end{figure}

Because of finite multiplicity fluctuations, the azimuthal angle $\Phi_{R}$ of the vector 
$\vec{Q}$ can differ from the azimuthal angle of the true reaction plane  $\phi_{true}$ by 
a deviation $\Delta\phi_{pl}=\phi_{true} -\Phi_R$. For corresponding 
coefficients $v_{1}^{true}$ and $v_{2}^{true}$ of Fourier expansion of azimuthal angle 
distribution of particles with respect to the "true" reaction plane then holds 
$v_{1}^{true}$= $v_{1}$/$\langle cos\Delta\phi_{pl} \rangle$
and 
$v_{2}^{true}$= $v_{2}$/$\langle cos2\Delta\phi_{pl} \rangle$. 

For the determination of these corrections  we follow
the procedure as detailed in ref. \cite{olit97,olit93}.
The distribution of $\Delta\phi_{pl}$ can be presented as a 
function of a single dimensionless parameter $\chi$ which measures the 
accuracy of the reaction-plane determination. For a large sample of events within
the same centrality window, according to the central limit theorem, the fluctuations
of the length $Q=\mid\vec{Q}\mid$ of the reaction-plane vector around its average value, 
$\langle Q \rangle$, are distributed as a Gaussian:
\begin{equation}
\frac{dN}{QdQd\Delta\phi_{pl}}=\frac{1}{\pi\sigma^{2}}exp\left[-\frac{|Q-\langle Q \rangle|^{2}}{\sigma^{2}}
\right]=\frac{1}{\pi\sigma^{2}}exp\left[-\frac{Q^{2}+\langle
Q\rangle^{2}-2Q\langle{Q}\rangle cos\Delta\phi_{pl}}{\sigma^{2}}\right],
\end{equation}
The fluctuations with standard deviation $\sigma$ are assumed to be isotropic since the 
azimuthal anisotropies are small. The dimensionless parameter $\chi$=$\langle Q\rangle/\sigma$
scales with the particle multiplicity $M$ like $\sqrt M$. Integration of Eq. 4 over both
$\Delta\phi_{pl}$ and $Q$ yields the value 
$\langle cos~n\Delta\phi_{pl} \rangle$, which will be used later for the correction
of the measured azimuthal anisotropies. According to the "Ollitrault method" we find:

\begin{equation}
\langle cos~n\Delta\phi_{pl} \rangle=\langle cos~n\Delta\phi_{pl}^{Ollit} \rangle=\frac{\sqrt{\pi}}{2}\chi
e^{-\chi^{2}/2}\left[
I_{\frac{n-1}{2}}\left(\frac{\chi^2}{2}\right)+I_{\frac{n+1}{2}}\left(\frac{\chi^2}{2}\right)\right]
\end{equation}
where $I_{k}$ is the modified Bessel function of the order k. 

Fig.~\ref{fg:bess} shows the variations of the first two correction 
coefficients for n=1 (correction for the directed flow signal) and n=2 (correction for the elliptic
flow signal)  with $\chi$. 

In order to determine the parameter $\chi$ we randomly divided 
the  hits in each event into two subgroups containing each one half of the number of
particles. For each of the two subevents one can construct according to Eq. 3 the two independent vectors
$\vec{Q_1}$ and $\vec{Q_2}$, respectively, and extract the angle
$\Delta \Phi_{1,2}$=$\Phi_{R,1}-\Phi_{R,2}$ between the two vectors.
Fig.~\ref{fg:rplane} shows the distribution of the 
relative azimuthal angle $\Delta \Phi_{1,2}$.

The ratio of events with $\mid\Delta \Phi_{1,2}\mid >90^{\circ}$ to the total number of
events allows to determine the parameter $\chi$ from Eq. 6:
\begin{equation}
\frac{N(90^{0}<\Delta\Phi_{1,2}<180^{0})}{N(0^{0}<\Delta\Phi_{1,2}<180^{0})}
=\frac{exp(-\chi^2/2)}{2}
\end{equation}

Alternatively, these corrections can be estimated by the adapted 
%\cite{org89,zhan90,htun99}
\cite{org89} -- \cite{htun99}
transverse-momentum method of Danielewicz and Odyniec \cite{trmom}. First, we determined the average of
the weighted unit vector of transverse velocity $\langle V_{x}^{meas}\rangle$ of each charged particle, 
projected on the reaction plane. The orientation of the reaction plane is estimated by the vector 
$\vec{Q}_{i}^{m}$, determined here by excluding the considered charged particle in order to remove 
autocorrelations:
\begin{equation} 
\langle V_{x}^{meas} \rangle=\overline{\left(\frac{\vec{r}_{i}}{\mid \vec{r}_{i}\mid}\right)_{i}
\frac{\vec{Q}_{i}^{m}}{\mid\vec{Q}_{i}^{m}\mid}},~~~
\vec{Q}_{i}^{m}=\sum^{M}_{j\neq i}Z_{j}\frac{\vec{r}_{j}}{\mid \vec{r}_{j}\mid}.
\end{equation} 
Then, we calculated the average of the unit vector of transverse velocity $\langle 
V_{x}^{true}\rangle$
projected on the "true" reaction plane,
\begin{equation} 
\langle V_{x}^{true} \rangle=\left[\frac{\overline{Q^{2}-M}}{\overline{M(M-1)}}\right]^{1/2},~~~
\vec{Q}=\sum^{M}_{i}\frac{\vec{r}_{i}}{\mid \vec{r}_{i}\mid}.
\end{equation} 
The corrections due to finite reaction-plane resolution 
for the directed-flow signal (n=1) and the elliptic-flow signal (n=2)
can be determined from the following equations:
\begin{equation} 
\langle cos\Delta\phi_{pl}^{Dan}\rangle
=\frac{\langle V_{x}^{meas}\rangle}{\langle V_{x}^{true}\rangle}, 
~~~~~\langle cos2\Delta\phi_{pl}^{Dan}\rangle=2\cdot\langle cos^2\Delta\phi_{pl}^{Dan}\rangle-1
\end{equation}

\begin{table}
\begin{center}
 \begin{tabular}{|c|c|c|c|c|c|} \hline
Reaction &\multicolumn{3}{|c|}{$^{58}$Ni+$^{58}$Ni at 1.9 A GeV}
&\multicolumn{2}{|c|}{$^{40}$Ca+$^{nat}$Ca at 2$A$ GeV}\\ \hline
 M$_{react}$ & 2 - 6 & 7 - 10 & $\ge$ 11 & 1 - 3 & $\ge$ 4\\ \hline
$\sigma_{pl}$  & 43$^{\circ}$  & 46$^{\circ}$ & 50$^{\circ}$ &  52$^{\circ}$ & 55$^{\circ}$\\ \hline
$\langle cos2\Delta\phi_{pl}^{Ollit} \rangle$  & 0.37  & 0.32 & 0.26 &  0.22  & 0.19 \\ \hline
\hline
$\langle cos2\Delta\phi_{pl}^{Dan} \rangle$  & 0.33  & 0.28 & 0.23 &  0.18  & 0.16 \\ \hline
 \end{tabular}
 \caption{ The width (standard deviation $\sigma_{pl}$) of the relative azimuthal-angle distributions, the 
reaction plane resolution for the elliptic-flow signal (n=2) determined by "Ollitrault method" 
($\langle cos2\Delta\phi_{pl}^{Ollit}\rangle$) 
and by the adapted transverse-momentum method ( $\langle cos2\Delta\phi_{pl}^{Dan}\rangle$) (see text)
for studied bins in hit multiplicity M$_{react}$ of the reaction detector.}
 \label{tb:res}
\end{center}
\end{table} 

Our further analysis was restricted to a sufficient  vector length 
$\mid\vec{Q}\mid > 3$ in order to reject the most 
central events lacking spectator flow. This selection rejects 25$\%$ of all registered events. 
The resulting values of the correction coefficient for the elliptic-flow signal due to finite 
reaction-plane resolution  
$\langle cos2\Delta\phi_{pl} \rangle$, determined by both methods, are given in
Table~\ref{tb:res} for each bin in hit multiplicity M$_{react}$ 
of the reaction detector for both systems studied. 
The systematic error of these values 
was estimated from the difference 
$|\langle cos2\Delta\phi_{pl}^{Olitt}\rangle - \langle cos2\Delta\phi_{pl}^{Dan}\rangle|$ 
and is found to be less than 15$\%$.
We also studied the influence of the cut $\mid\vec{Q}\mid > 3$ on these values. The removal
of the cut leads to decrease in the correction value, see left panel of
Fig.~\ref{fg:rplane}. and 
Eq.6. 
But simultaneously also deduced values of $v_{2}$ decrease 
and hence the $v_{2}^{true}$ values
remain unchanged.

For comparison with previous studies we calculated also reaction plane resolution 
$\sigma_{pl}=\sigma_{1,2}/2$, where standard deviation $\sigma_{1,2}$  
is extracted from a Gaussian fit to the $dN/d\Delta\Phi_{1,2}$
distributions. 
Reaction plane resolution varies between $43^{\circ}$ and $55^{\circ}$ 
depending on the reaction centrality and the 
colliding system (see Table~\ref{tb:res}). These values agree with published data from studies of charged-baryon flow in 
similar colliding systems \cite{trmom,ritman95}. Moreover, the   
the  directed flow of charged baryons in the target-like rapidity region 
was observed for both colliding systems (see Appendix A). 
This verification demonstrates the quality of the reaction-plane determination. 

\subsection{Photon-particle discrimination}

\begin{figure}
  \vspace{-0.5cm}
  \begin{center}
    \mbox{
     \epsfxsize=9.5cm
     \epsffile{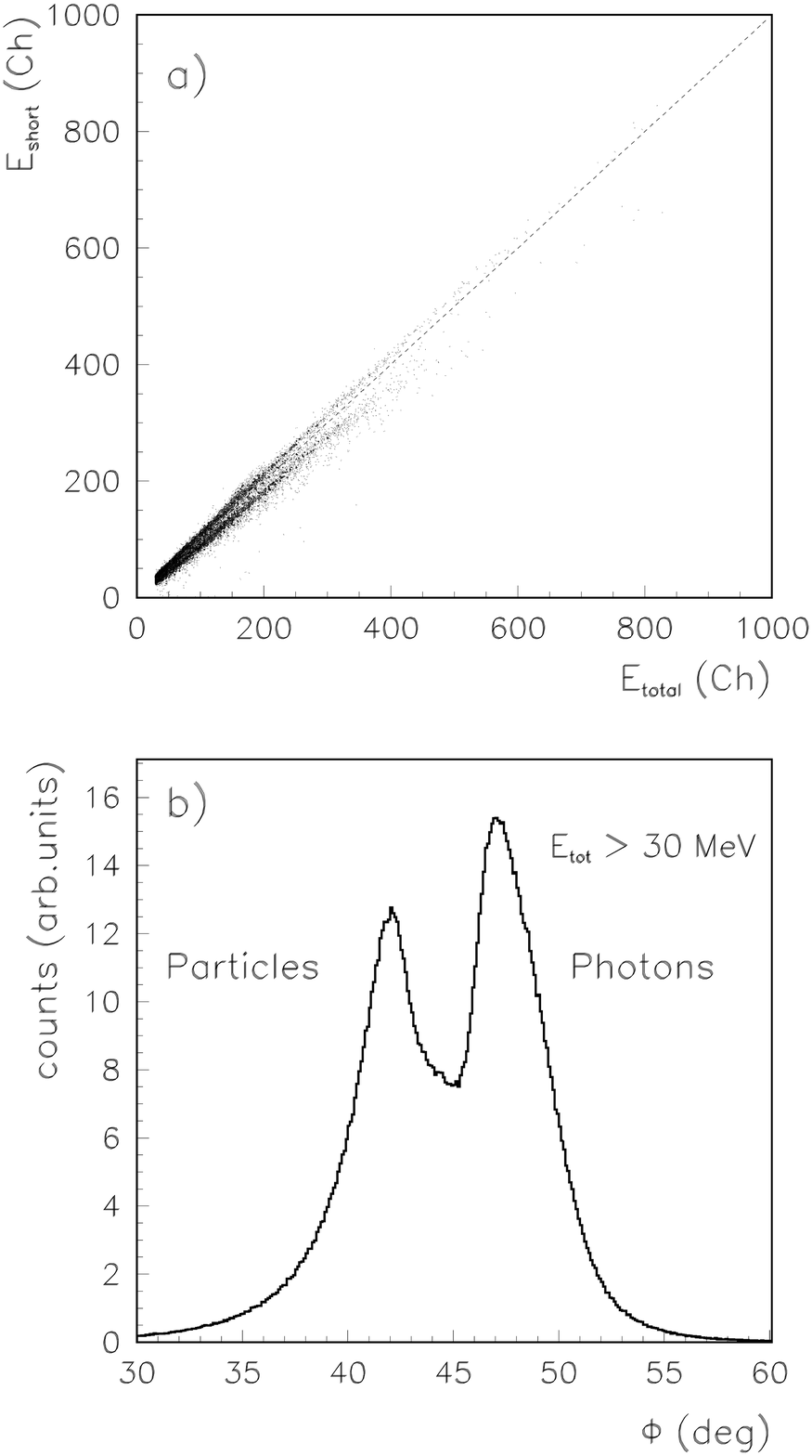}
      }
  \end{center}
  \vspace{0.0cm}
  \begin{center}\parbox{14.6cm}{
     \caption{Pulse-shape analysis of BaF$_2$ signals: 
a) graph of E$_{short}$ versus E$_{total}$ for the $\pi^{0}$ trigger in Ni+Ni reactions at 1.9 A GeV;
b) distribution of the pulse-shape parameter $\Phi_{PSA}$ for all 384 BaF$_2$ modules. In this case the 
pulse-shape analysis was performed only for E$_{total}>$30 MeV.}
     \label{fg:psa}
  }\end{center}
  \vspace{0.0cm}
\end{figure}

Efficient photon-particle discrimination in TAPS can be achieved by the combination of three methods 
%\cite{berg,fmarq,wagn,ralf97}:
\cite{berg} -- \cite{ralf97}:
\begin{description}
\item[1)]
The anticoincidence between the BaF$_2$ and the corresponding CPV
was required to discriminate charged particle hits. 
\item[2)]
 The time-of-flight analysis allowed to separate photons
 from  massive particles (with non-relativistic velocities). 
\item[3)] 
The pulse-shape of the BaF$_2$ energy signal was analyzed. The scintillation light
in the BaF$_2$ crystal exhibits two components with different time constants $\tau$, one fast
($\tau$=0.6 ns) and one slow ($\tau$=620 ns). The intensity ratio of fast to slow component
depends on the ionization density of the incident radiation: it is larger for leptons (electrons following
photon interactions) than for other charged particles (p,d,t,$\alpha$...). The pulse-shape (PSA) parameter 
is defined as the angle $\Phi_{PSA}$ = $\arctan(E_{short}/E_{total})$, where E$_{short}$ is the fraction of
energy obtained by integrating the analog signal within the first 30 ns, and E$_{total}$ is the total energy 
obtained by integrating the analog signal within 2$\mu$s. Fig.~\ref{fg:psa}a presents a two-dimensional graph of
E$_{short}$ versus E$_{total}$ for all BaF$_2$ modules, and the distribution of the corresponding 
PSA parameter is shown in Fig.~\ref{fg:psa}b. The separation between photons and charged particles is obvious.

\end{description}

\subsection{Shower reconstruction by cluster analysis}

Hadrons, like protons and charged mesons, lose their energy by excitation and ionization. On the other hand,
high energy photons impinging on TAPS crystals generate an electromagnetic shower which fires several 
neighbouring detector modules. Therefore, the reconstruction algorithm to separate photons and hadrons and to
recover position and energy of the incident photon is based on defining a shower as a continuous cluster of 
individual responding detectors \cite{fmarq,ralf97,tap94a,tap94b}. 
In the first step, the fired detector modules are sorted according to the deposited energy. The combination to
clusters starts with the detector which indicates the maximum energy deposit in a block of crystals above a 
certain energy threshold E$_{high}$ (typically $\simeq$20 MeV). This so-called central detector is combined with
all directly neighbouring detectors (maximum 6 modules in the first ring of neighbours). However, these modules 
are only accepted in the cluster if the time difference between the neighbouring and the central module is within
the allowed time window and an energy deposition above the lower threshold E$_{low}$ (typically $\simeq$3 MeV) is
encountered. If the sum energy of this minimal cluster exceeds a threshold of typically 400-500 MeV or if the
fraction of the energy deposited in the central module is lower than 60-70 $\%$ of the sum energy, then the 
second ring of neighbouring detectors (up to 12 modules) is inspected for addition to the existing cluster. 
All allocated detectors are then removed from the detector hit pattern and the cluster search continues with 
the remaining highest-energy module. This procedure limits the cluster 
size to a maximum of 19 detector modules which is, however, sufficient to reconstruct even the highest energy 
photons encountered in this experiment.

In the second step the characteristics of all modules within a cluster were
compared to distinguish different types of showers. The cluster is called a photon-like cluster if:
\begin{description} 
\item[1.] the hit in the central detector is neutral, i.e. no CPV signal was found in the corresponding CPV module
or the CPV's of the neighbouring detectors.
\item[2.] the pulse shape corresponds to the hit of a photon (see
Fig.~\ref{fg:psa}). 
\item[3.] the time-of-flight is within the prompt time window. 
The neighbouring energy deposition is added only if the neighbours also have photon-like characteristics. 
\end{description}

The average particle occupancy in TAPS reaches values of 7$\%$ per module. This leads to
a finite probability that the clusters in the TAPS block are connected or partly overlapping.
However, in general connected or overlapping clusters have a second maximum and there will
be more energy deposited in the second ring as is expected from a single hit. 
The clusters with a second maximum were removed from  the further analysis.
The energy of the cluster $E$ results from the sum of energies $E_i$ deposited
in the individual detectors, and the direction of the incident photon  $\vec{r}$
is reconstructed from an energy-weighted sum over the vectors $\vec{r_i}$ pointing
from the target to each responding detector from the cluster \cite{awe92}.
\begin{equation}
        E=\sum_{i}E_{i}~~~~~and~~~~~~
	\vec{r} =\frac{\sum_{i}W_{i}\vec{r_i}}{\sum_{i}W_{i}}~~~
	where~~~W_{i}=MAX\left(0,\left[W_{0}+ln\frac{E_{i}}{E}\right]\right).
\end{equation}
Parameter $W_0$ controls the importance of the positions of the neighbouring
modules with respect to the position of the center module.

\subsection{Neutral-meson reconstruction}

\begin{figure}[h]
  \vspace{-1.0cm}
  \begin{center}
    \mbox{
     \epsfxsize=10.0cm
     \epsffile{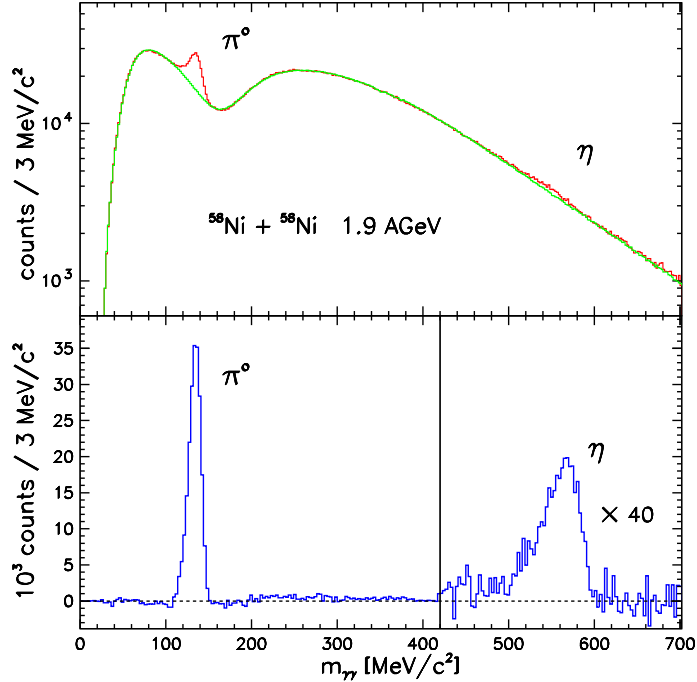}
      }
  \end{center}
  \vspace{-0.65cm}
  \begin{center}\parbox{13.7cm}{
     \caption{Results of the invariant-mass analysis of photon pairs.
 The upper frame shows the invariant-mass
spectrum which corresponds to the $\eta$ trigger in the experiment $^{58}$Ni+$^{58}$Ni at 1.9 AGeV.
The combinatorial background (dotted line) was determined by event mixing. The lower frame
shows the invariant-mass distribution after background subtraction and demonstrates the quality
of the background determination.}
    \label{fg:ima_nini}
  }\end{center}
  \vspace{-0.45cm}
\end{figure}

\begin{figure}
  \vspace{-2.0cm}
  \begin{center}
    \mbox{
     \epsfxsize=8.0cm
     \epsffile{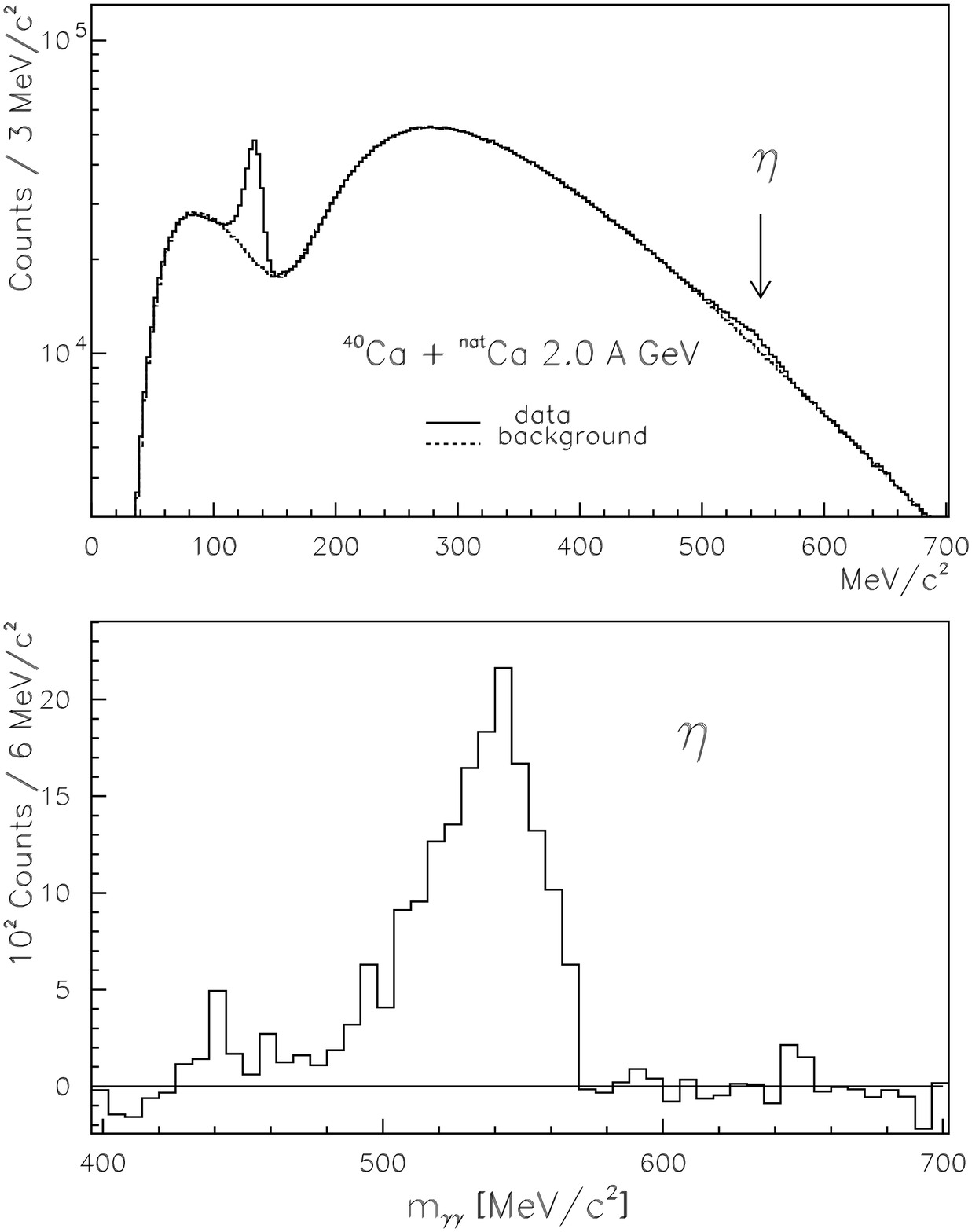}
      }
  \end{center}
  \vspace{-1.0cm}
  \begin{center}\parbox{14.4cm}{
     \caption{Results of the invariant-mass analysis of photon pairs.
 The upper frame shows the invariant-mass
spectrum which corresponds to the $\eta$ trigger in the experiment $^{40}$Ca+$^{nat}$Ca at 2.0 AGeV.
The combinatorial background (dotted line) was determined by event mixing. The lower frame
shows the invariant-mass distribution after background subtraction.}
     \label{fg:eima_caca}
   }\end{center}
   \vspace{-0.45cm}
\end{figure}

\begin{figure}
  \vspace{-1.0cm}
  \begin{center}
    \mbox{
     \epsfxsize=8.0cm
     \epsffile{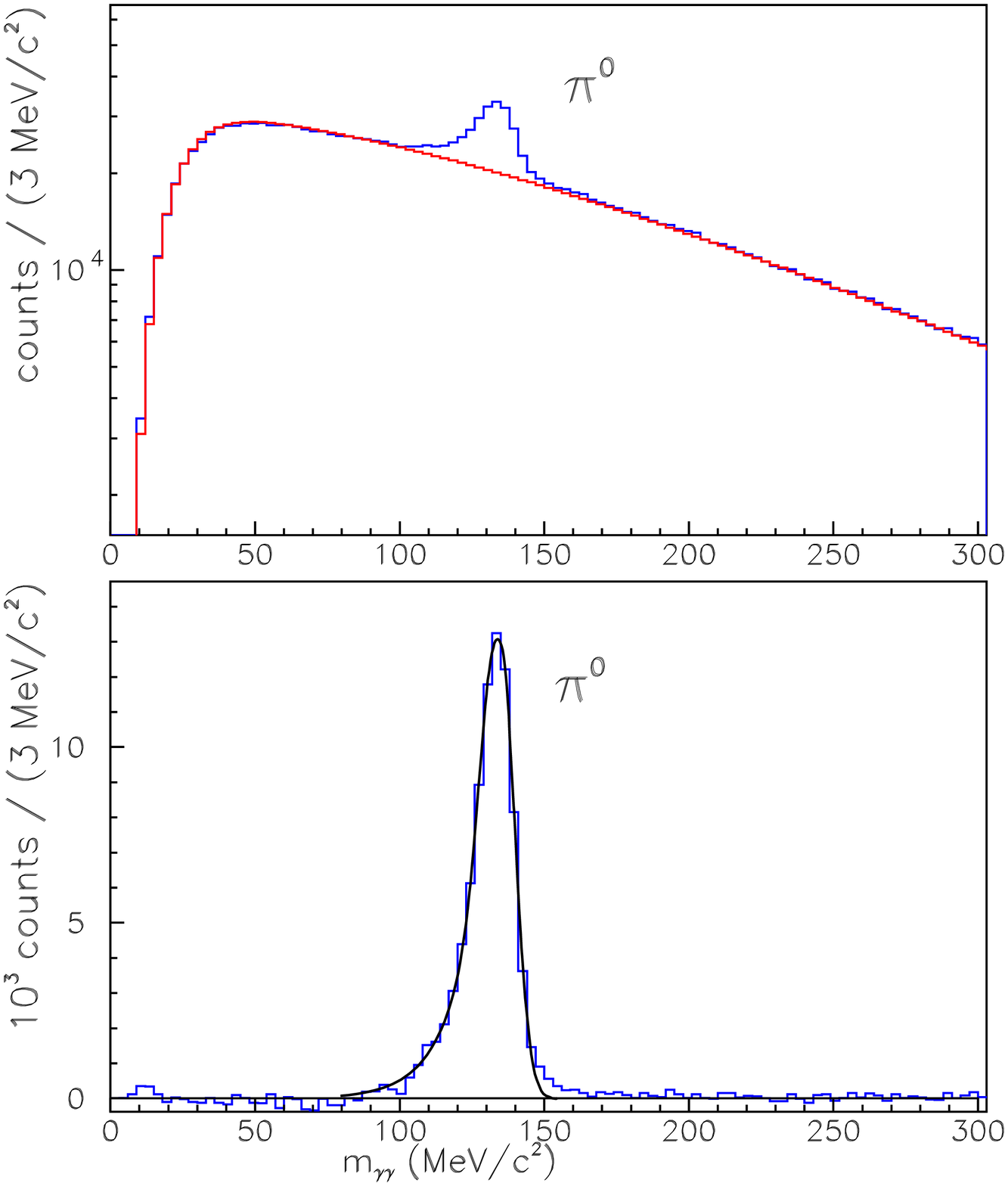}
      }
  \end{center}
  \vspace{-1.0cm}
  \begin{center}\parbox{14.4cm}{
     \caption{Results of the invariant-mass analysis of photon pairs.
 The upper frame shows the invariant-mass
spectrum which corresponds to the $\pi^{0}$ trigger in the experiment $^{40}$Ca+$^{nat}$Ca at 2.0 AGeV.
The combinatorial background (dotted line) was determined by event mixing. The lower frame
shows the invariant-mass distribution after background subtraction.}
     \label{fg:pima_caca}
  }\end{center}
  \vspace{-0.45cm}
\end{figure}

Neutral pions and $\eta$-mesons can be detected by the measurement of their two-photon decay,
 which still carries the signature of the meson in its invariant mass.
For each pair of detected photons in a given event we calculated the
invariant mass $M_{pair}$ and the momenta $\vec{p}_{pair}$
using the following relations:
\begin{equation} 
M_{pair}^2=2E_{1}E_{2}(1-cos\Theta_{12})~~~
and~~~
\vec{p}_{pair}=\vec{p}_{1}+\vec{p}_{2} ,
\end{equation} 
\noindent where $E_{1}$, $\vec{p}_{1}$ and
$E_{2}$, $\vec{p}_{2}$ are the  energies and momenta of the corresponding photons, and
$\Theta_{12}$ is the opening angle of the photon pair. We analysed 
only neutral mesons in a narrow rapidity window ($|y-y_{cm}|\le$ 0.1). 
The resulting invariant-mass distributions 
(Fig.~\ref{fg:ima_nini} -- \ref{fg:pima_caca}) reveal peaks centered at values 
of the $\pi^{0}$- and $\eta$-meson rest mass. These peaks are located on top of a large combinatorial
background which originates from uncorrelated photon pairs. 

The high particle multiplicity in heavy-ion collisions, limited acceptance and the detector efficiency
usually lead to very low signal-to-background ratios, especially at lower 
transverse momenta. Therefore, the precise knowledge of the shape and magnitude of the combinatorial
background is required for a reliable evaluation of the neutral-meson peak contents. The analysis quality 
can be improved by increasing the purity of the photon sample, i.e.,
by reducing the combinatorial background. Therefore, all cuts in the pulse-shape analysis 
and cluster reconstruction scheme were optimized for a maximum signal-to-background ratio in the
invariant-mass spectrum. The resulting signal-to-background ratios for the studied bins in 
reaction centrality M$_{react}$ and transverse momentum p$_t$ are listed in the
Table~\ref{tb:SB} for 
both reactions studied. 
If the background has a monotonous falling or rising shape it will not
be difficult to approximate the shape of the background with an analytic function. However,
usually the artificial structure in the invariant-mass spectra created by limited detector
acceptance could not be reproduced by any analytical fit of the background.

\begin{table}
\begin{center}
 \begin{tabular}{|c|c|c|c|c|c|} \hline
Reaction &\multicolumn{3}{|c|}{$^{58}$Ni+$^{58}$Ni at 1.9 A GeV} &\multicolumn{2}{|c|}{$^{40}$Ca+$^{nat}$Ca at 2$A$ GeV}\\
 M$_{react}$ & 2 - 6 & 7 - 10 & $\ge$ 11 & 1 - 3 & $\ge$ 4\\ \hline
~~~~~~~~~~~~~~~~p$_t$ (MeV/c) & & & & & \\
$\eta$~~~~~~~~~~0 - 600 &3.9$\%$ & 3.1$\%$ & 2.5$\%$ & 8.9$\%$   &   4.5$\%$ \\ \hline
$\pi^{0}$~~~~~~~~~0 - 200   &   9.5$\%$   & 6.2$\%$   & 4.3$\%$   &15$\%$  &10$\%$  \\
~~~~~~~~~~~~200-400         &   21$\%$  & 13$\%$  & 9.2$\%$   &36$\%$  &24$\%$  \\
~~~~~~~~~~~~400-600         &   61$\%$  & 33$\%$  & 24$\%$  &99$\%$  &65$\%$  \\
~~~~~~~~~~~~600-800         &   83$\%$  & 47$\%$  & 36$\%$  &140$\%$ &110$\%$   \\ \hline
 \end{tabular}
 \caption{The signal-to-background ratio (S/B) for the $\pi^0$ and $\eta$  reconstruction for different bins
in multiplicity M$_{react}$ and transverse momentum p$_t$. }
 \label{tb:SB}
\end{center}
\end{table}

\subsection{ Method of event mixing}

Since the combinatorial background consists of pairs of uncorrelated particles, the natural way
of its reconstruction is the use of the so-called "event-mixing" technique
by combining photons randomly selected from different events 
%\cite{zajc84,fok92,lho94}.
\cite{zajc84} -- \cite{lho94}.
This method is based on the assumption that photons from different events are not correlated. Therefore, 
the mass signals
( $\pi^{0}$, $\eta$-mesons) are absent when we construct a two-photon invariant mass distribution
by taking photons from different events. This procedure automatically takes into
account the detector acceptance and the detection efficiency. Such combinations can be formed in large numbers,
so that the combinatorial background can be determined with high statistical accuracy. Then the background 
distribution is normalized outside of the meson-mass window to the measured invariant-mass distribution and 
subtracted from the latter. The meson-peak contents can be extracted from the background-free invariant-mass 
spectra by integrating within the meson-mass window.

\begin{figure}
  \vspace{-2.9cm}
  \begin{center}
    \mbox{
     \epsfxsize=13.0cm
     \epsffile{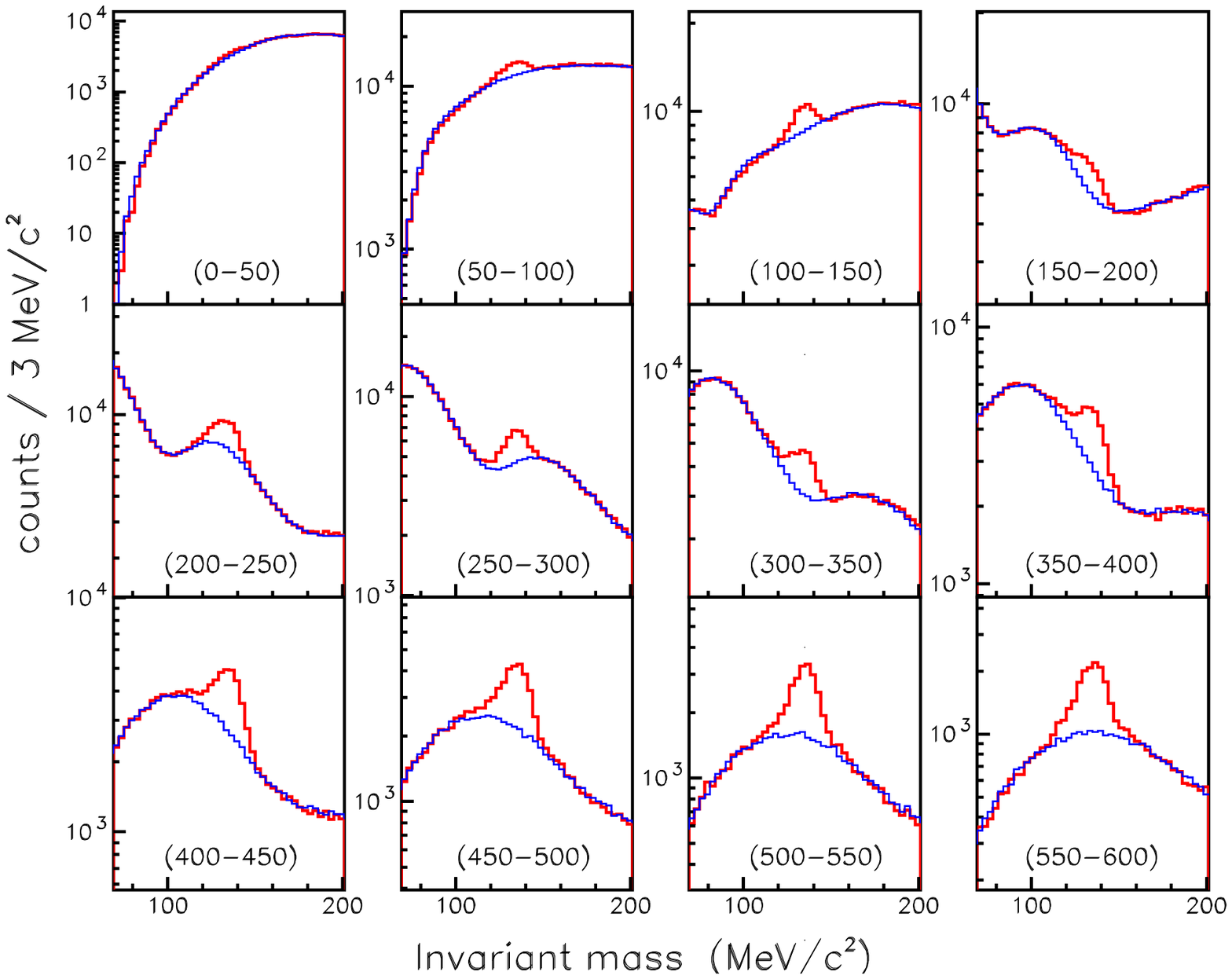}
      }
  \end{center}
  \vspace{-4.4cm}
  \begin{center}\parbox{13.7cm}{
    \caption{Invariant-mass spectra corresponding to the $\pi^{0}$ trigger
 in the experiment $^{58}$Ni+$^{58}$Ni at 1.9 AGeV for different bins in
 transverse momentum p$_t$ (range in MeV/c is noted in each spectrum)
 and photon multiplicity M$_{\gamma}$=2. The combinatorial background was determined by event mixing
 and is shown as a dashed line together with the histogram for the experimental data in each case.}
    \label{fg:ima_pt}
  }\end{center}
  \vspace{-0.6cm}
\end{figure}

However, for a perfect description of the combinatorial background special care has to be taken to ensure that
the data and background distributions are generated for the same class of events with very
similar event characteristics. Therefore, only photons from the same event class may be mixed.
An event class is defined by the analysed trigger TRG and the photon 
multiplicity M$_{\gamma}$ found in the event. Including the  constraint on M$_{\gamma}$ in the event
classification is important as the signal-to-background ratio (S/B) depends on M$_{\gamma}$ \cite{awes1}.
The phase-space distribution of the mixed photon pairs must correspond to the measured ones.
The phase space covered by TAPS allows the identification of neutral mesons emitted in
a narrow window around midrapidity only. As a result, the transverse momentum of
the photon pair can classify the phase space allocation. Therefore, if one divides
the photon pairs in the classes according to TRG and M$_{\gamma}$ into groups according to their transverse 
momenta p$_t$, an identical phase space allocation for mixed and measured pairs can be expected. 
As an example, Fig.~\ref{fg:ima_pt} shows the invariant-mass distributions for different
p$_t$-classes and the photon-multiplicity class M$_{\gamma}$=2 ($\pi^{0}$
trigger for the experiment Ni+Ni at 1.9 A GeV).
The normalized combinatorial background is shown as a dashed line together with the histogram for the
experimental data in each case. The spectral form of the distributions varies 
strongly with the transverse momentum which reflects the opening angle acceptance of TAPS.

\begin{figure}
  \vspace{-1.0cm}
  \begin{center}
    \mbox{
     \epsfxsize=8.0cm
     \epsffile{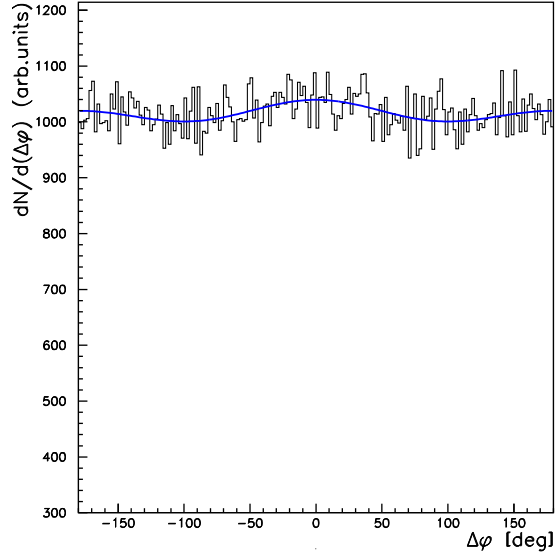}
      }
  \end{center}
  \vspace{-1.0cm}
  \begin{center}\parbox{14.2cm}{
     \caption{Azimuthal-angle distributions of inclusive
photons detected by TAPS from peripheral (2$\le M_{react} \le$6) Ni+Ni reactions at 1.9 AGeV, averaged 
over photon transverse momentum. The solid line is a result of the fit by the first two terms of the Fourier 
expansion \cite{vol96}. }
     \label{fg:photon}
  }\end{center}
  \vspace{0.0cm}
\end{figure}

\begin{figure}
  \vspace{-1.7cm}
  \begin{center}
    \mbox{
     \epsfxsize=14.0cm
     \epsffile{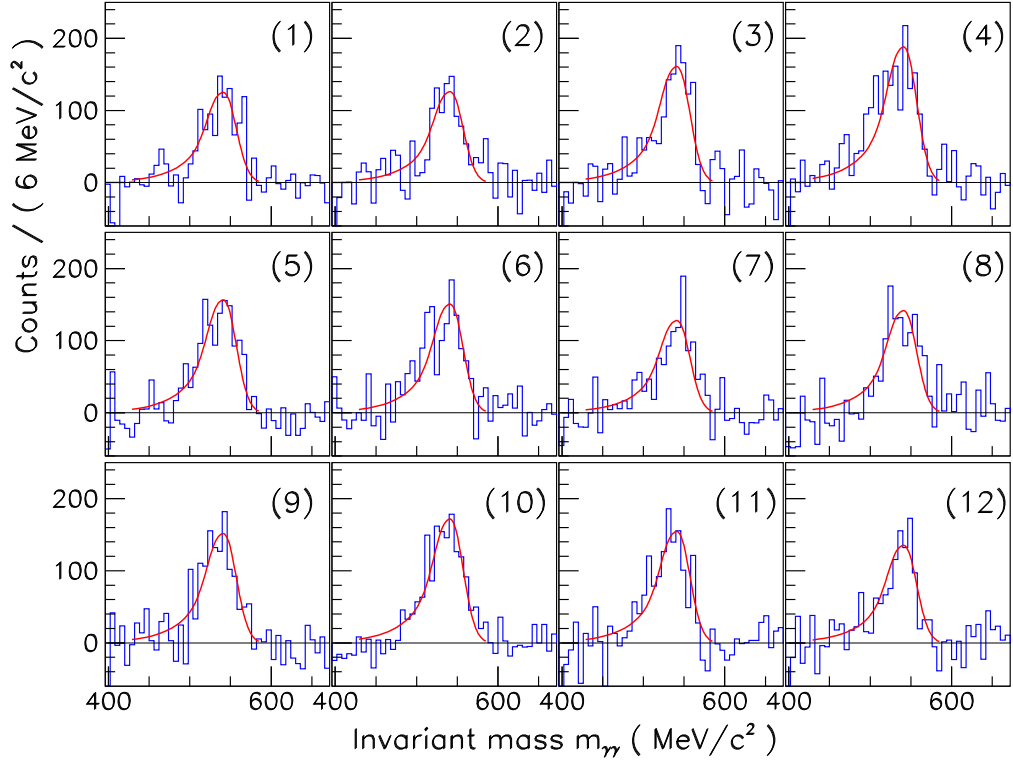}
      }
  \end{center}
  \vspace{-1.0cm}
  \begin{center}\parbox{14.2cm}{
    \caption{The extraction of the $\eta$-meson peak contents in the interval 0$\le$p$_t\le$ 400 MeV/c in
the 12 equal intervals in azimuthal angle $\Delta\varphi=\phi_{pair}-\Phi_R$ 
of meson emission relative to the reaction plane $\Phi_R$. The fitted line is an asymmetric
Gaussian distribution \cite{mat90}. The number of $\eta$ mesons for each bin in azimuthal
angle $\phi$ was obtained by integrating the background-free invariant-mass
distribution within the meson-mass window: 470 MeV/c$^{2}\le$m$_{\gamma\gamma}\le$ 570
MeV/c$^{2}$. The reaction is $^{40}$Ca+$^{nat}$Ca at 2 AGeV.}
    \label{fg:ima_phi}
  }\end{center}
  \vspace{0.0cm}
\end{figure}

\begin{figure}
  \vspace{-1.0cm}
  \begin{center}
    \mbox{
     \epsfxsize=10.0cm
     \epsffile{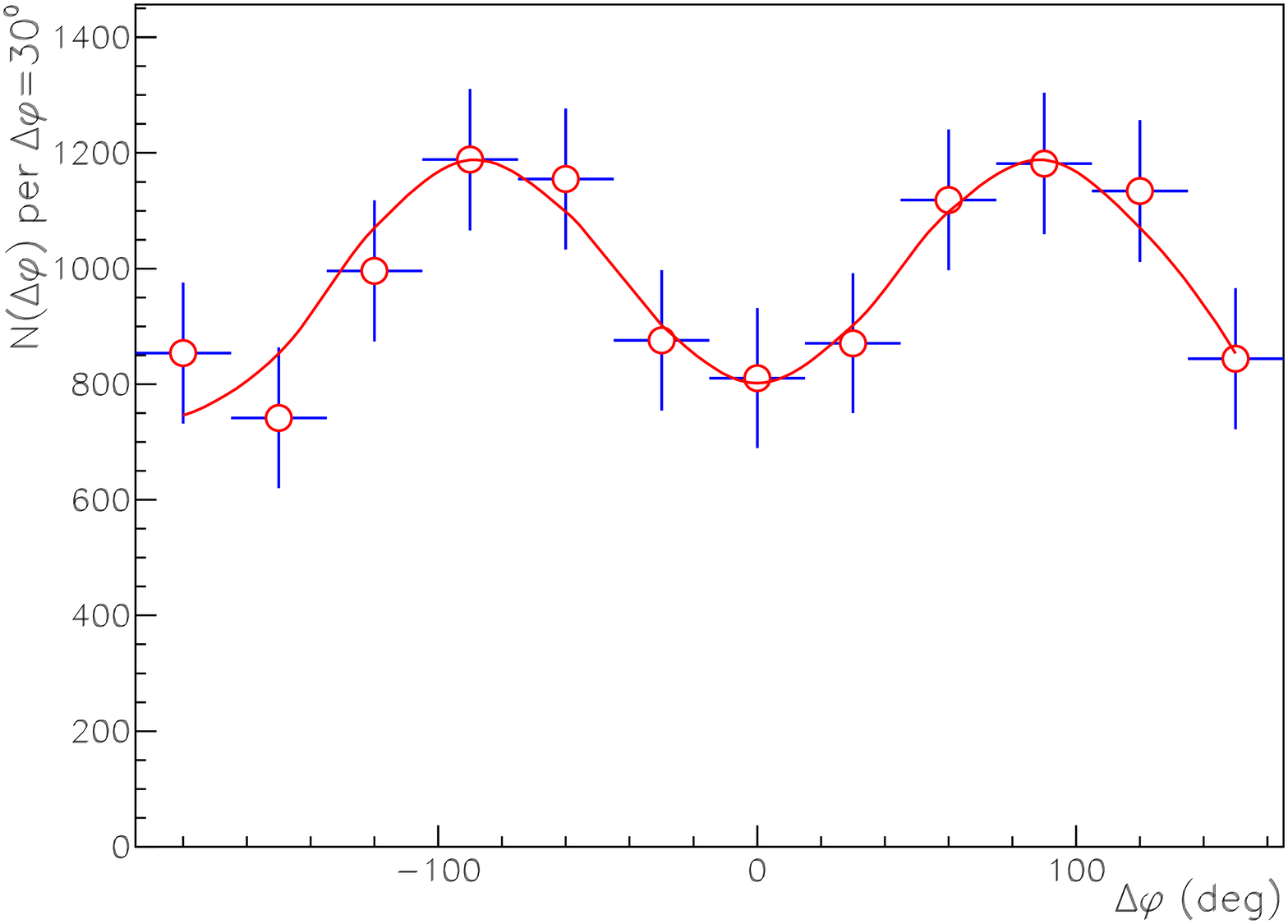}
      }
  \end{center}
  \vspace{-1.0cm}
  \begin{center}\parbox{14.2cm}{
    \caption{Azimuthal-angle distribution of $\eta$ mesons emitted in 
$^{40}$Ca+$^{nat}$Ca collisions at 2.0 AGeV  averaged over multiplicity M$_{react}$ and transverse
momentum p$_t$. The solid line is the result of a fit by the first two terms of the 
Fourier expansion. }
    \label{fg:etaflow}
  }\end{center}
  \vspace{0.0cm}
\end{figure}

Subsequently, the mixed photon pairs were selected to have an opening-angle distribution close to 
the opening angle of the original pair. This selection can be implemented by an additional classification
according to the TAPS block combination.
However, the initial concept of a good reproduction of the phase-space occupancy of the
invariant-mass distribution in the combinatorial background by the method of event mixing
can still suffer in some cases. Besides the true correlations due to the neutral-meson decay 
there are some additional correlations due to reaction dynamics or detector effects. These
correlations are also destroyed by event mixing and therefore are not removed from the data.

In a mixed event two photons can be very close to each other in space. In a real event, if
two photons are too close to each other, they are not resolved, and are counted as one photon.
This effect was reduced by storing photons only when they belong to different detector blocks 
as in the experiment. This 
leads to a small under-estimation of the background from mixing events, but only for very
small invariant masses (below 50 MeV). 
The photon pairs from meson-peak regions ($\pi^{0}$,$\eta$) will be more abundant 
than photon pairs from the regions outside of it \cite{zajc84}. Therefore, if the
combinatorial background will be generated from all photon pairs from the meson-peak regions,
we will overestimate the contribution from these photons. The ratio of the yield of the meson
peak to the yield of the total invariant mass  S/T=Signal/(Signal+Background) was used to
suppress photon pairs from the resonance regions for each event class (TRG, M$_{\gamma}$, p$_t$).

We observed weak azimuthal anisotropy found in the yield of inclusive
photons detected by TAPS.  
As an example, Fig.~\ref{fg:photon} shows the azimuthal angle distributions of inclusive
photons detected by TAPS with respect to the reaction plane from peripheral (2$\le M_{react} \le$6) 
Ni+Ni reactions at 1.9 A GeV averaged over photon transverse momentum. The Fourier analysis of this distribution shows the presence of weak 
positive elliptic flow of inclusive photons $v_{2}\simeq .01$, indicating a 
preferred emission of photons in the reaction plane. Furthermore, the detailed analysis revealed that 
the magnitude of the combinatorial background is also weakly dependent 
($v^{bg}_{2}\simeq .01$)
on the azimuthal angle $\Delta\varphi=\phi_{pair}-\Phi_R$
of meson emission relative to the reaction plane.
As the main source of photons in our case is due to $\pi^{0}$ decay, 
both these anisotropies can be attributed
to the anisotropy in the pion emission. As the integrated pion yields 
are dominated by pions with low transverse momentum p$_t$, the elliptic flow of low
p$_t$ neutral pions is also expected to be weak and positive. 
The performed Monte-Carlo simulation confirmed this assumption. The simulation also 
shows that 
the correction for this effect can be done by normalizing the combinatorial 
background for each 
bin in azimuthal angle $\Delta\varphi$ separately. 

In our analysis we used 12 equal 
intervals in azimuthal angle $\Delta\varphi=\phi_{pair}-\Phi_R$ 
of meson emission relative to the reaction plane $\Phi_R$. The normalization factor was
found to be a smooth function of $\Delta\varphi$. 

In order to extract the azimuthal-angle distributions of mesons 
the following algorithm was used: 
In a first step, the calculated invariant-mass distributions for each studied trigger and reaction-centrality 
window were divided into several sections according to the photon multiplicity $M_{\gamma}$, 
transverse momenta p$_t$, TAPS block combination and azimuthal angle $\Delta\varphi$.
Then the combinatorial background was generated by the outlined scheme of event mixing,
normalized outside of the meson-mass windows and subtracted from the invariant-mass
spectra for each section separately. Then the background-free invariant-mass spectra from each section were 
summed. As an example, Fig.~\ref{fg:ima_phi} shows the resulting background-free 
invariant mass spectra in the $\eta$ meson mass region for 12 equal 
intervals in $\Delta\varphi$ summed over the reaction centrality
 M$_{react}$ and transverse momentum p$_t$. The $\eta$ meson peak content
 was obtained by integrating over the mass window 
 470 MeV/c$^2\le$M$_{\gamma\gamma}\le$570 MeV/c$^2$, and for the $\pi^{0}$ mesons we used
the mass window 100 MeV/c$^2\le$M$_{\gamma\gamma}\le$150 MeV/c$^2$.
The resulting azimuthal-angle distribution of $\eta$ mesons is shown in
Fig.~\ref{fg:etaflow}. 

The uncertainty in the meson-yield extraction is determined by the statistical errors
in the contents of the total mass distribution in between the integration limits and the number
of entries in the (scaled) combinatorial background distribution. The combined statistical
error is taken to be the square root of the quadratic sum of these contributions. \\

\subsection{ Simulation}

\begin{figure}
  \vspace{-0.6cm}
  \begin{center}
    \mbox{
     \epsfxsize=10.2cm
     \epsffile{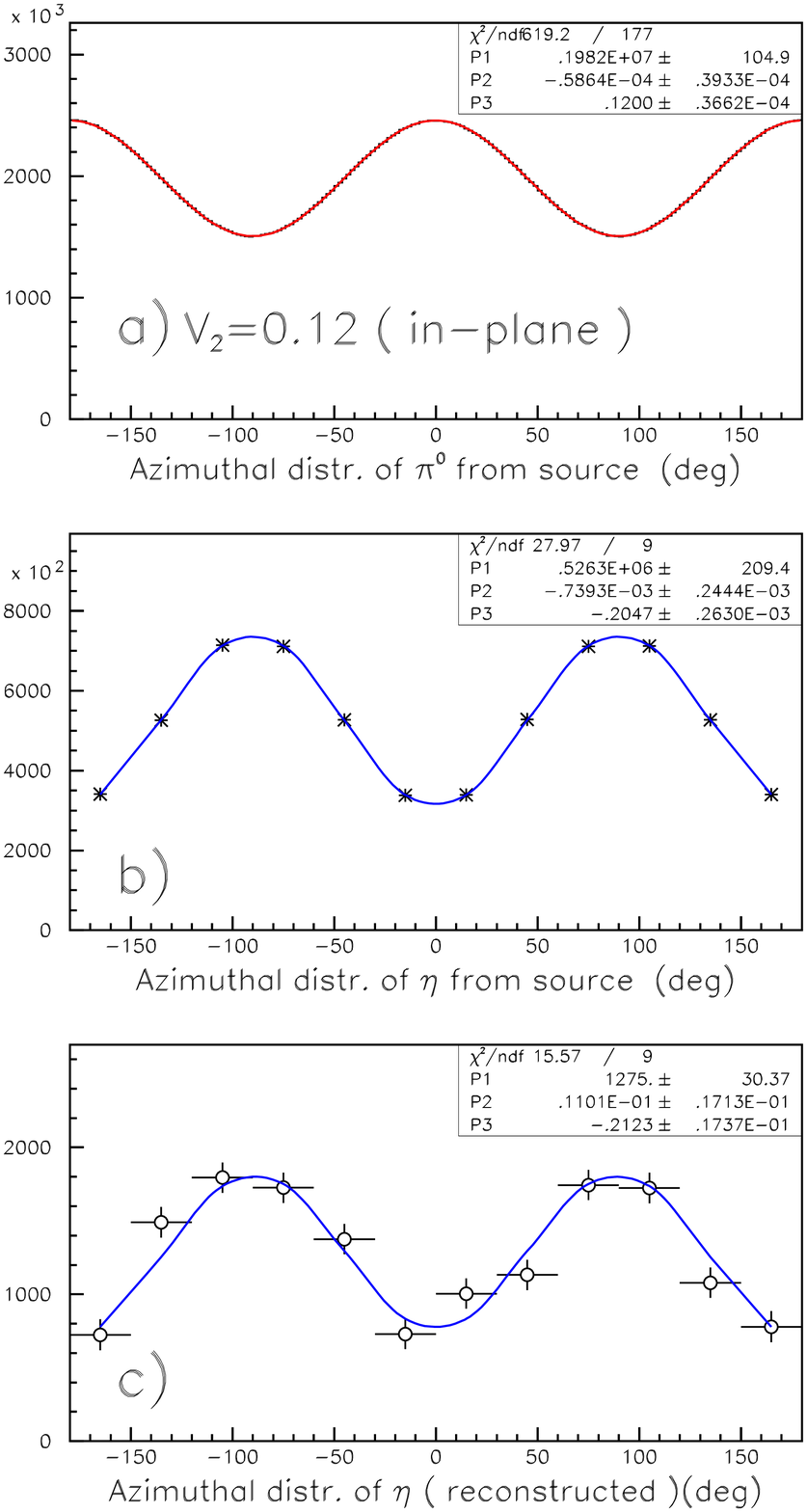}
      }
  \end{center}
  \vspace{-1.0cm}
  \begin{center}\parbox{14.4cm}{
     \caption{Azimuthal-angle distributions from a Monte-Carlo simulation, a) for $\pi^{0}$
from a pion source with strong {\underline positive} elliptic flow; b) for $\eta$ mesons from the $\eta$ source;
c) for $\eta$ mesons reconstructed by the proposed scheme of event mixing.}
     \label{fg:sim_plus}
  }\end{center}
  \vspace{-0.5cm}
\end{figure}
 
\begin{figure}
  \vspace{-0.6cm}
  \begin{center}
    \mbox{
     \epsfxsize=11.2cm
     \epsffile{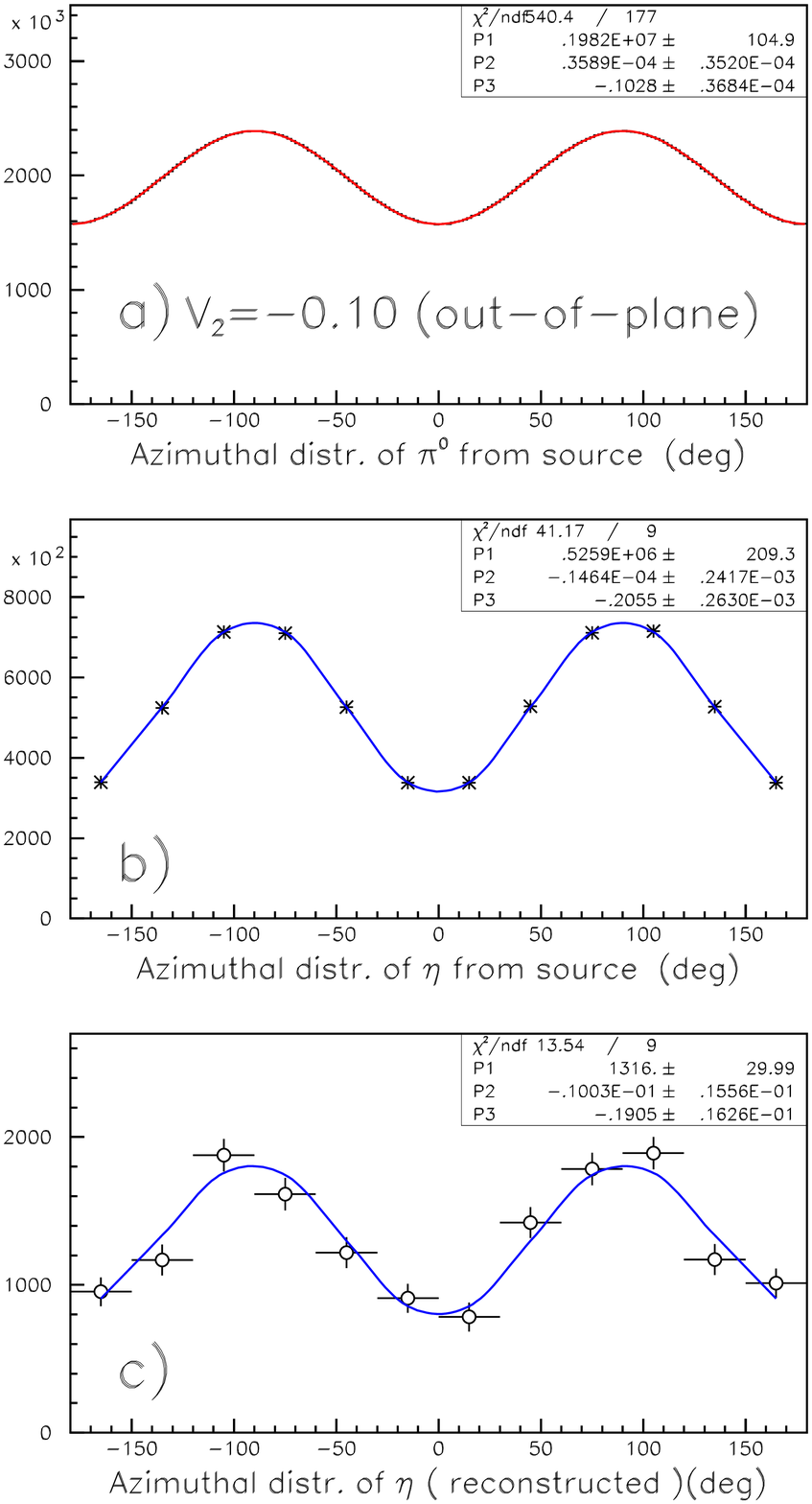}
      }
  \end{center}
  \vspace{-1.0cm}
  \begin{center}\parbox{14.4cm}{
     \caption{Azimuthal angle distributions from a Monte-Carlo simulation, a) for $\pi^{0}$
from a pion source with a strong {\underline negative} elliptic flow; b) for $\eta$ mesons from the $\eta$ source;  
c) for $\eta$ mesons reconstructed by the proposed scheme of event mixing.}
     \label{fg:sim_neg}
  }\end{center}
  \vspace{-0.5cm}
\end{figure} 

As the main source of combinatorial background in our case are uncorrelated photon pairs
which mainly originate from $\pi^{0}$ decay, the weak anisotropy of the background 
can be attributed to the anisotropy in pion emission.
To test this assumption and to check the accuracy of the 
proposed procedure of event mixing a Monte-Carlo simulation was performed in
which the emission of $\pi^{0}$ and $\eta$ mesons is governed by a thermal distribution
\begin{equation}
\frac{dN}{dE} \propto const \cdot x \cdot \sqrt{x^{2}-1}\cdot Ee^{(-E/T)}
\end{equation}
where  $E$ and $x=E/M$ are the total energy of the meson ( M is the meson mass )
and its reduced total energy in the center-of-mass system, respectively. The meson ( $\pi^{0}$, $\eta$ )
source parameters, i.e. temperature $T$ and 
meson production probability per participant were chosen consistently with those measured \cite{paul}.
The polar-angle distribution of the emitted mesons is isotropic in the center-of-mass system. 

The  azimuthal-angle
distribution $ N(\Delta\varphi$) of the emitted mesons with respect to the reaction
plane is not isotropic and is given by :
\begin{equation}
  N( \Delta\varphi ) = const \cdot ( 1 + 2\cdot v_{2}^{true}\cdot cos(2\Delta\varphi) ),
\end{equation}
where the parameter $v_{2}^{true}$ reflects the strength of elliptic flow. 
The anisotropy in the emission of $\eta$ mesons was 
assumed to be $v_{2}^{true}$=-0.2, close to the experimental value.
The simulations were performed with several values of the
anisotropy in the emission of pions ($-0.12<v_{2}^{true}<+0.12$), 
see below range of values  
actually observed in the experiment. 
Geometrical acceptance, on-line trigger schemes  and 
realistic detector resolution effects (time-of-flight and energy resolution of BaF$_2$ modules)
were taken into account. We used the identical procedure of analysis for the simulated data as for the 
experimental ones. The statistics in the simulated data was chosen close to that of the experimental data.
The results of the simulation show that
\begin{itemize}
\item the azimuthal anisotropy of the combinatorial background is an indication of the collective
flow of pions, and its sign and magnitude are determined by the anisotropy of the pionic flow.
Isotropy in the emission of pions leads to an azimuthally isotropic combinatorial background.

\item the extracted values for the elliptic-flow signal of $\eta$ and $\pi^{0}$ 
mesons are, within statistical errors, comparable with  source parameters; the proposed method
of event mixing works correctly in case of different anisotropies of the pion source.
As an example we present the results of such simulations 
for a pion source with strong positive (see Fig.~\ref{fg:sim_plus}) 
and negative elliptic flow (see Fig.~\ref{fg:sim_neg})
\end{itemize}

Based on these simulation we estimate a systematical 
error in the parameter $v_{2}$ for the experimental data, 
which are influenced by much smaller mean pion anisotropy and 
hence also combinatorial background anisotropy $v^{bg}_2 \simeq +0.01$, 
to be below 2$\%$.

\section{Results}

\begin{figure}
  \vspace{-2.9cm}
  \begin{center}
    \mbox{
     \epsfxsize=11.6cm
     \epsffile{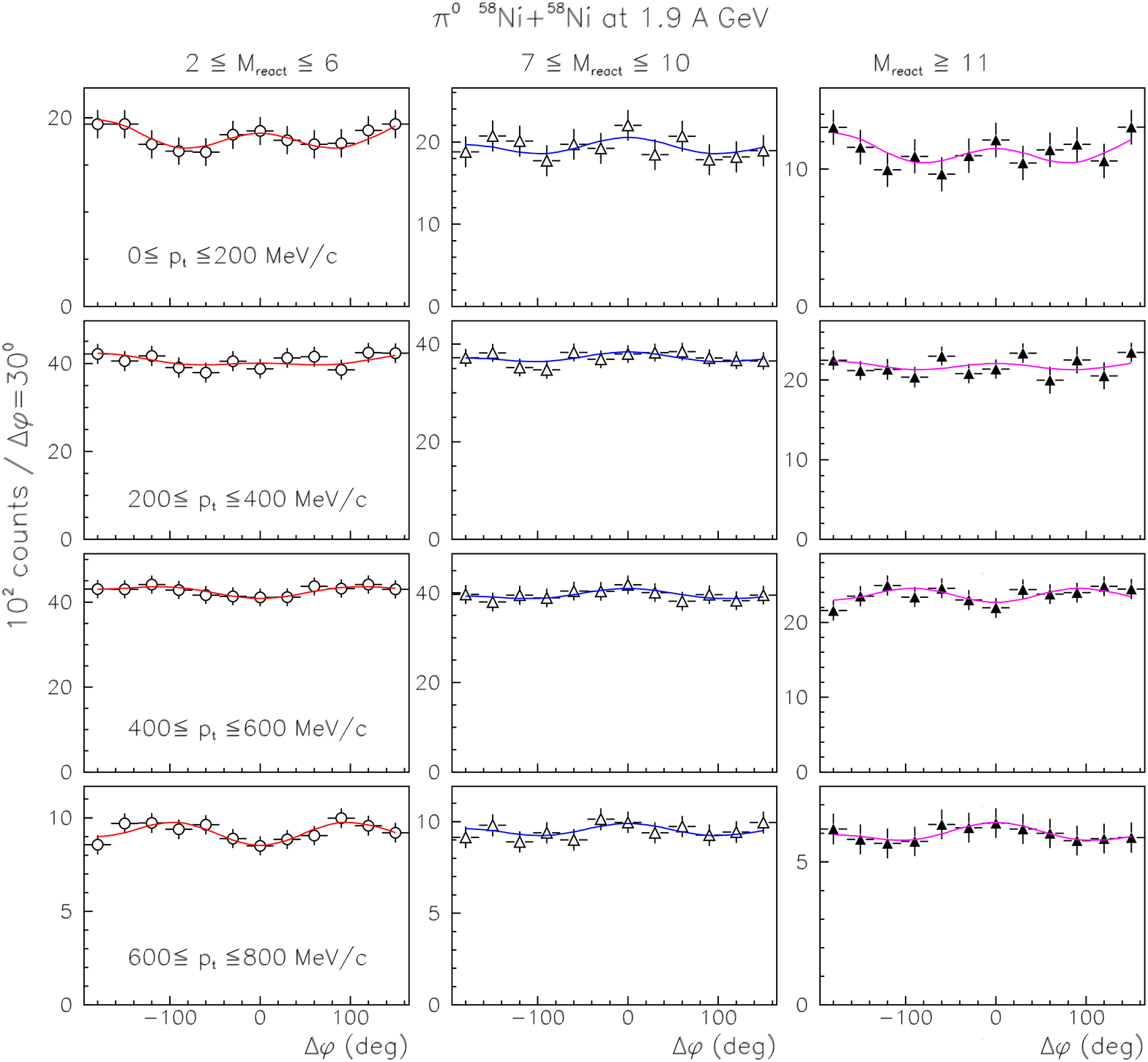}
      }
  \end{center}
  \vspace{-0.9cm}
  \begin{center}\parbox{14.0cm}{
     \caption{ Azimuthal-angle distribution of $\pi^0$ mesons 
     emitted in $^{58}$Ni+$^{58}$Ni reactions at
     1.9~AGeV.
The panels correspond to different bins in reaction centrality
M$_{react}$ and transverse momenta p$_t$.}
     \label{fg:pflo_nini}
  }\end{center}
  \vspace{0.0cm}
\end{figure}

\begin{figure}
  \vspace{-1.3cm}
  \begin{center}
    \mbox{
     \epsfxsize=8.9cm
     \epsffile{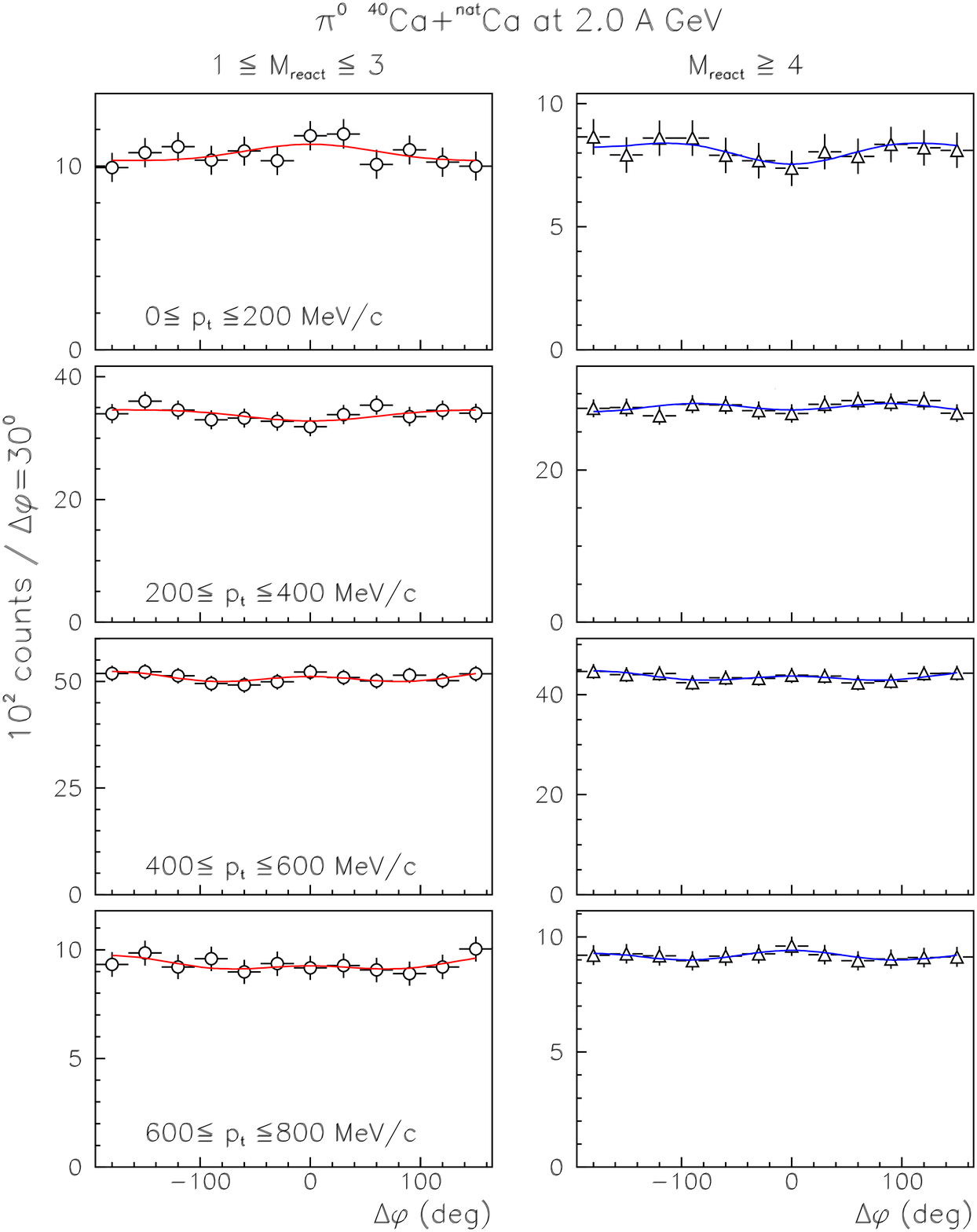}
      }
  \end{center}
  \vspace{-0.9cm}
  \begin{center}\parbox{14.0cm}{
     \caption{ Azimuthal angle distribution of $\pi^0$ mesons emitted in 
$^{40}$Ca+$^{40}$Ca reactions at 2.0~AGeV. The panels correspond
to different bins in reaction centrality M$_{react}$ and transverse momenta p$_t$. }
     \label{fg:pflo_caca}
  }\end{center}
  \vspace{0.0cm}
\end{figure}

\begin{figure}
  \vspace{-4.0cm}
  \begin{center}
    \mbox{
     \epsfxsize=15.0cm
     \epsffile{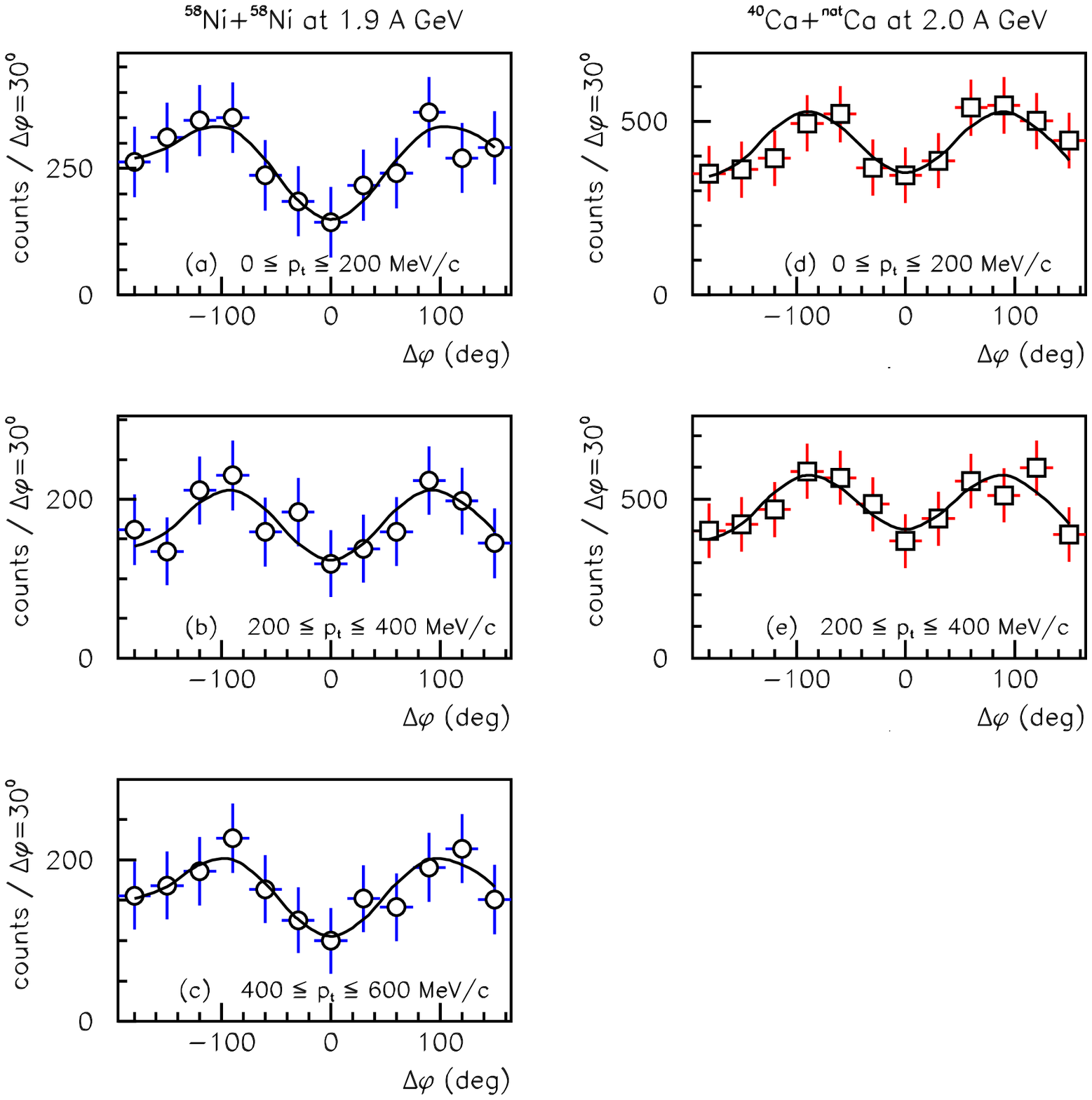}
      }
  \end{center}
  \vspace{-1.0cm}
  \begin{center}\parbox{14.3cm}{
     \caption{Azimuthal-angle distribution of $\eta$ mesons with respect
to the reaction plane for different bins in transverse momentum p$_t$ and
averaged over multiplicity M$_{react}$: 
(a) - (c) for the experiment  $^{58}$Ni+$^{58}$Ni at 1.9~AGeV and (d) - (e) for the
experiment $^{40}$Ca+$^{nat}$Ca at 2~AGeV.}
     \label{fg:eflo_pt}
  }\end{center}
  \vspace{-1.0cm}
\end{figure} 

\begin{figure}
  \vspace{-1.0cm}
  \begin{center}
    \mbox{
     \epsfxsize=15.7cm
     \epsffile{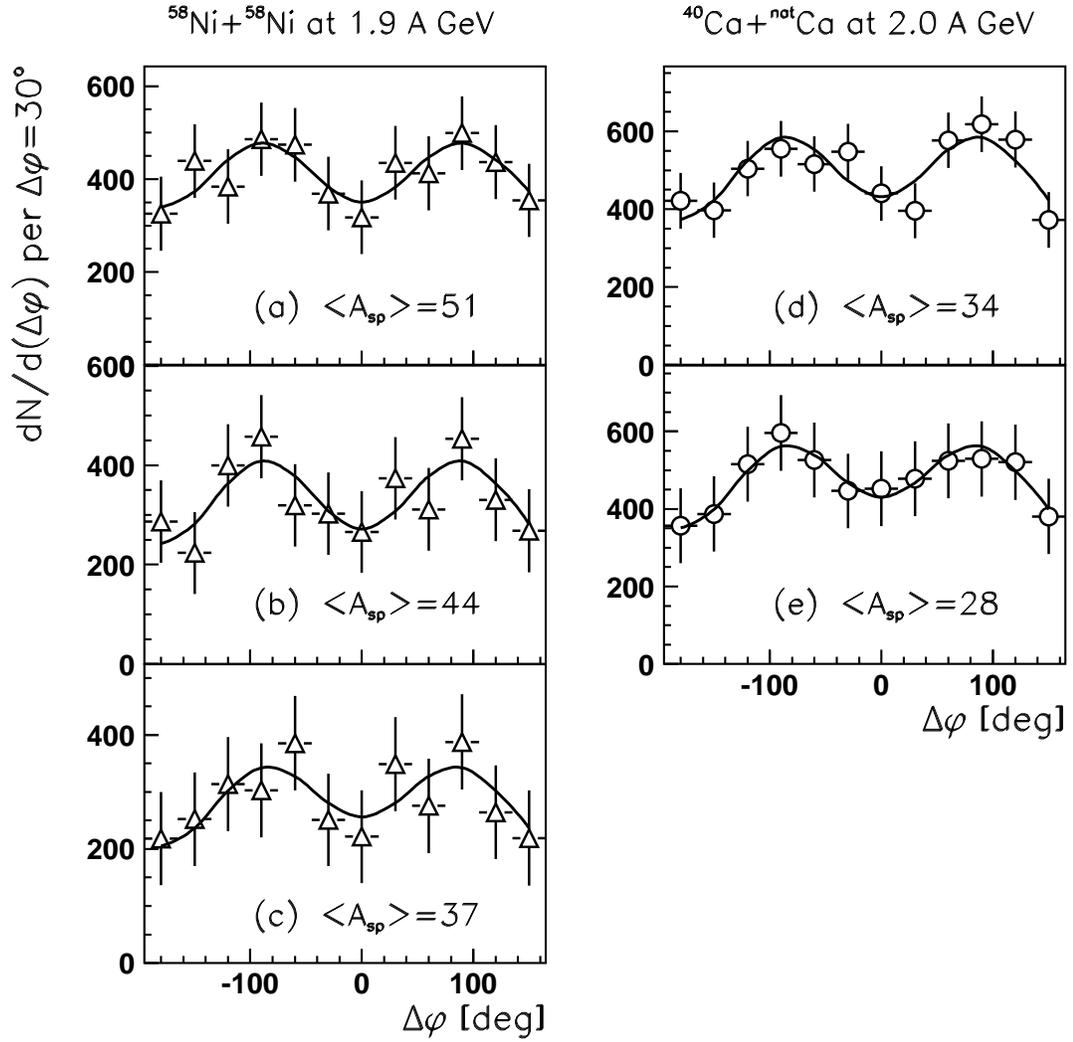}}
  \end{center}
  \vspace{-1.0cm}
  \begin{center}\parbox{15.0cm}{
     \caption{ 
Azimuthal-angle distribution of $\eta$ mesons with respect
to the reaction plane for different bins in multiplicity M$_{react}$ (reaction centrality)
(a) - (c) for the experiment  $^{58}$Ni+$^{58}$Ni at 1.9~AGeV and (d) - (e) for the
experiment $^{40}$Ca+$^{40}$Ca at 2~AGeV, see Table~\ref{tb:Asp}. The corresponding values of the
mean number of projectile-like spectator nucleons $\langle A_{sp}\rangle$ are indicated.}
     \label{fg:eflow_mul}
  }\end{center}
  \vspace{0.0cm}
\end{figure}

We  present in Fig.~\ref{fg:pflo_nini} -- \ref{fg:eflow_mul} the resulting azimuthal yields of $\eta$ and $\pi^{0}$ mesons for 
different bins in M$_{react}$ and transverse momentum p$_t$ for both  systems studied. 
Because of low statistics for $\eta$ mesons we present the p$_t$ dependence (see Fig. 20) without a
selection on the reaction centrality M$_{react}$, 
and the reaction centrality dependence without a selection on the transverse momentum p$_t$ (see Fig. 21).
We fitted the azimuthal yields of $\eta$ and $\pi^0$  mesons by the first two terms of a 
Fourier expansion in the azimuthal angle:
\vskip 0.2cm
\begin{equation}
N(\Delta\varphi)=\frac{N_0}{2\pi}\pmatrix{1+2v_1 cos(\Delta\varphi) + 2v_2 cos(2\Delta\varphi)}.
\end{equation}
The first coefficient ($v_1$) is used to parametrize the in-plane emission of the particles
parallel ($v_1>0$) or antiparallel ($v_1<0$) to the impact-parameter vector
(directed flow), whereas the second coefficient ($v_2$) quantifies an elliptic flow, which is
negative for out-of-plane emission and positive for an in-plane emission \cite{olit,vol96}. 
The extracted values of $v_1$ are zero within the error bars, as should be expected  
since we study symmetric colliding systems at midrapidity, see
Tables~\ref{tb:v12pt}--\ref{tb:v1}. 
The resulting values of the parameter $v_2$ for $\eta$ and $\pi^0$  mesons 
are given in Table~\ref{tb:v12pt} and \ref{tb:v2}.

\begin{table}
\begin{center}
 \begin{tabular}{|c|c|c|c|c|c|} \hline
Reaction &\multicolumn{3}{|c|}{$^{58}$Ni+$^{58}$Ni at 1.9 AGeV}
&\multicolumn{2}{|c|}{$^{40}$Ca+$^{nat}$Ca at 2~AGeV}\\ 
p$_t$ (MeV/c) & 0-200 & 200-400 & 400-600 & 0-200 & 200-400\\ \hline
$v_1$ & -0.08$\pm$0.07 & -0.03$\pm$0.05 & -0.07$\pm$0.08 & 0.01$\pm$0.04 & 0.02$\pm$0.04\\ \hline
$v_2$ & 0.10$\pm$0.06 & 0.11$\pm$0.05 & 0.10$\pm$0.05 & 0.11$\pm$0.04 &0.10$\pm$0.04 \\ \hline
 \end{tabular}
 \caption{Parameters $v_1$ and $v_2$ (not corrected for the reaction-plane resolution)
 for $\eta$ mesons deduced from the experimental azimuthal angle
distributions (see Fig.~\ref{fg:eflo_pt}) for several intervals in transverse momentum p$_t$, 
without a selection on the reaction centrality.}
 \label{tb:v12pt}
\end{center}
\end{table} 

\begin{table}
\begin{center}
 \begin{tabular}{|c|c|c|c|c|c|} \hline
Reaction &\multicolumn{3}{|c|}{$^{58}$Ni+$^{58}$Ni at 1.9~AGeV}
&\multicolumn{2}{|c|}{$^{40}$Ca+$^{nat}$Ca at 2~AGeV}\\ 
M$_{react}$ & 2 - 6 & 7 - 10 & $\ge$ 11 & 1 - 3 & $\ge$ 4\\ \hline
 p$_t$ (MeV/c) &   \multicolumn{5}{|c|}{ $v_1$ for $\eta$ mesons}\\ \hline
0-600  & 0.01$\pm$0.06   & 0.06$\pm$0.10  & 0.11$\pm$0.16n &  0.05$\pm$0.06 & -0.02$\pm$0.08 \\ \hline
            & \multicolumn{5}{|c|}{$v_1$ for $\pi^0$ mesons}\\ \hline
0 - 200     & -0.021$\pm$0.026 & 0.011$\pm$0.022 &-0.031$\pm$0.042   & -0.021$\pm$0.022  &  0.021$\pm$0.022 \\
200-400     & -0.016$\pm$0.017 & 0.016$\pm$0.017 & -0.011$\pm$0.021  & 0.006$\pm$0.011   &  -0.011$\pm$0.012 \\
400-600     & -0.017$\pm$0.032 & 0.010$\pm$0.022 & 0.006$\pm$0.017   & -0.005$\pm$0.011  &  -0.005$\pm$0.011 \\
600-800     & -0.012$\pm$0.021 & 0.016$\pm$0.017 & 0.016$\pm$0.032   &  0.006$\pm$0.012  &  -0.016$\pm$0.021 \\
\hline
 \end{tabular}
 \caption{Parameters $v_1$ (not corrected for the reaction-plane resolution)
for $\pi^{0}$ and $\eta$ mesons deduced from the experimental azimuthal 
distributions for several intervals in reaction centrality.}
 \label{tb:v1}
\end{center}
\end{table} 

\begin{table}
\begin{center}
 \begin{tabular}{|c|c|c|c|c|c|} \hline
Reaction &\multicolumn{3}{|c|}{$^{58}$Ni+$^{58}$Ni at 1.9~AGeV}
&\multicolumn{2}{|c|}{$^{40}$Ca+$^{nat}$Ca at 2~AGeV}\\
M$_{react}$ & 2 - 6 & 7 - 10 & $\ge$ 11 & 1 - 3 & $\ge$ 4\\ \hline
$\langle A_{sp} \rangle$ & 51 & 44 & 37 & 34 & 28  \\ \hline
p$_t$ (MeV/c) &\multicolumn{5}{|c|}{ $v_2$ for $\eta$ mesons}\\ \hline
0-600         & -0.08$\pm$0.04 & -0.12$\pm$0.05 & -0.10$\pm$0.06 &  -0.09$\pm$0.03 & -0.09$\pm$0.04 \\ \hline
              & \multicolumn{5}{|c|}{$v_2$ for $\pi^0$ mesons}\\ \hline
0 - 200   &   0.033$\pm$0.017  & 0.021$\pm$0.019  & 0.026$\pm$0.031 &  0.012$\pm$0.016  & -0.006$\pm$0.021 \\
200-400   &   0.005$\pm$0.007  & 0.007$\pm$0.009  & 0.011$\pm$0.016 & -0.011$\pm$0.006  & -0.005$\pm$0.011 \\
400-600   &  -0.004$\pm$0.005  & 0.008$\pm$0.011  &-0.018$\pm$0.016 &  0.006$\pm$0.007  &  0.016$\pm$0.007 \\
600-800   &  -0.031$\pm$0.012  & 0.012$\pm$0.013  & 0.022$\pm$0.021 &  0.005$\pm$0.012  &  0.007$\pm$0.012 
\\ \hline
 \end{tabular}
 \caption{Parameters $v_2$ (not corrected for the reaction-plane resolution)
for $\pi^{0}$ and $\eta$ mesons deduced from the experimental azimuthal 
distributions for several intervals (a-e) in M$_{react}$. The corresponding 
mean number of projectile-like spectators $\langle A_{sp} \rangle$ are indicated (see text). 
For $\pi^0$ mesons the selected intervals in transverse momentum p$_t$ are indicated as well.}
 \label{tb:v2}
\end{center}
\end{table} 

The obtained $v_2$ coefficients were corrected for the reaction-plane resolution as 
$v_{2}^{true}$= $v_{2}$/$\langle cos2\Delta\phi_{pl} \rangle$, where 
$\Delta\phi_{pl}=\phi_{true} -\Phi_R$ is the fluctuation of 
the azimuthal angle of the reconstructed reaction plane with 
respect to the true one, see sect. 3.2. 
The resulting dependence of $v_{2}^{true}$ as a function of transverse momentum p$_t$
is shown in Fig.~\ref{fg:v2tru_ept} -- \ref{fg:v2tru_ppt} for $\eta$ and $\pi^0$ mesons for both the Ni+Ni and Ca+Ca collisions. 
Note that the error-bars shown in Fig.~\ref{fg:v2tru_ept} -- \ref{fg:v2tru_ppt}
 represent statistical errors only.
The systematical errors in the resulting parameter $v_{2}^{true}$
are dominated by two sources: 

i) the uncertainty in the determination of the reaction-plane resolution 
$\langle cos2\Delta\phi_{pl} \rangle$ ( see sect. 3.2), which leads to 
a relative error in $v_{2}^{true}$ of the order of 15$\%$;

ii) the uncertainty induced by the variation of the azimuthal anisotropy of the combinatorial background 
with $v_{2}^{bg} \simeq 0.01$ which was estimated from a Monte-Carlo simulation 
and leads to relative errors in $v_{2}^{true}$
of approximately 2$\%$. 

Other sources of systematical errors like detector non-uniformities 
and occupancy of the TAPS spectrometer are found to be even smaller.

\section{Discussion}

The data shows strong negative elliptic flow of $\eta$ mesons ( $v_{2}^{true}<0$ )
for all studied bins in transverse momentum p$_t$ and
reaction centrality, indicating a preferred emission of $\eta$ mesons perpendicular to the
reaction plane. 
In contrast, the $v_{2}^{true}$ values for $\pi^{0}$ are 
close to zero within the error bars except for  
peripheral Ni+Ni collisions ($\langle A_{sp}\rangle=51$), where a weak elliptic-flow signal 
is observed, changing from positive to 
negative sign with increasing pion transverse momentum p$_t$, see
Fig.~\ref{fg:v2tru_ppt}.

\begin{figure}
  \vspace{.0cm}
  \begin{center}
    \mbox{
     \epsfxsize=8.0cm
     \epsffile{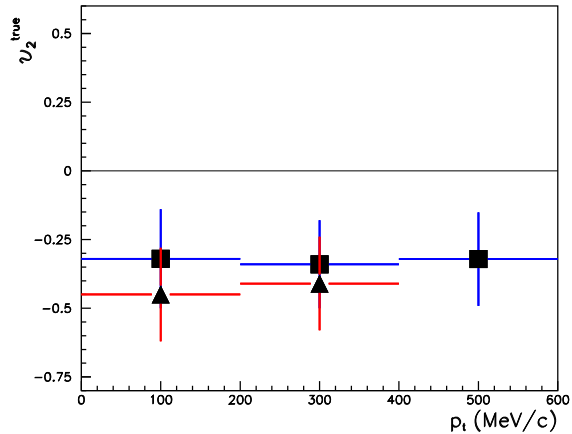}
      }
  \end{center}
  \vspace{-0.95cm}
  \begin{center}\parbox{14.4cm}{
     \caption{The parameters $v_{2}^{true}$ for $\eta$ mesons 
as a function of transverse momentum p$_t$ (averaged over multiplicity M$_{react}$):
boxes indicate results from the experiment  $^{58}$Ni+$^{58}$Ni at 1.9~AGeV and triangles the results from 
the experiment $^{40}$Ca+$^{nat}$Ca at 2~AGeV.}
     \label{fg:v2tru_ept}
  }\end{center}
  \vspace{-0.45cm}
\end{figure}

\begin{figure}
  \vspace{-1.0cm}
  \begin{center}
    \mbox{
     \epsfxsize=8.0cm
     \epsffile{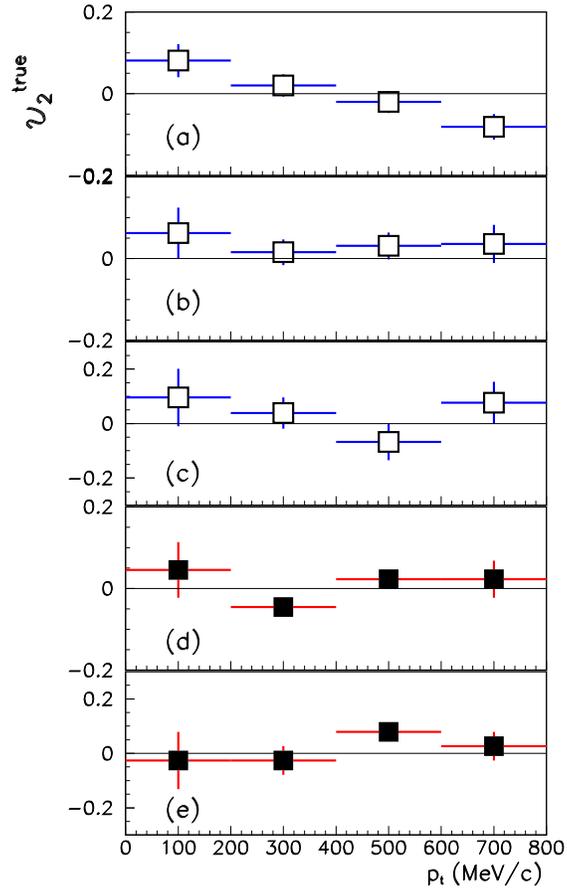}
      }
  \end{center}
  \vspace{-0.95cm}
  \begin{center}\parbox{14.4cm}{
     \caption{The parameters $v_{2}^{true}$ for $\pi^{0}$ mesons 
as a function of transverse momentum p$_t$ for different bins in multiplicity M$_{react}$:
(a) - (c) for the experiment  $^{58}$Ni+$^{58}$Ni at 1.9~AGeV and (d) - (e) for the
experiment $^{40}$Ca+$^{nat}$Ca at 2~AGeV.}
     \label{fg:v2tru_ppt}
  }\end{center}
  \vspace{-0.45cm}
\end{figure}

\begin{figure}
  \vspace{-1.0cm}
  \begin{center}
    \mbox{
     \epsfxsize=13cm
     \epsffile{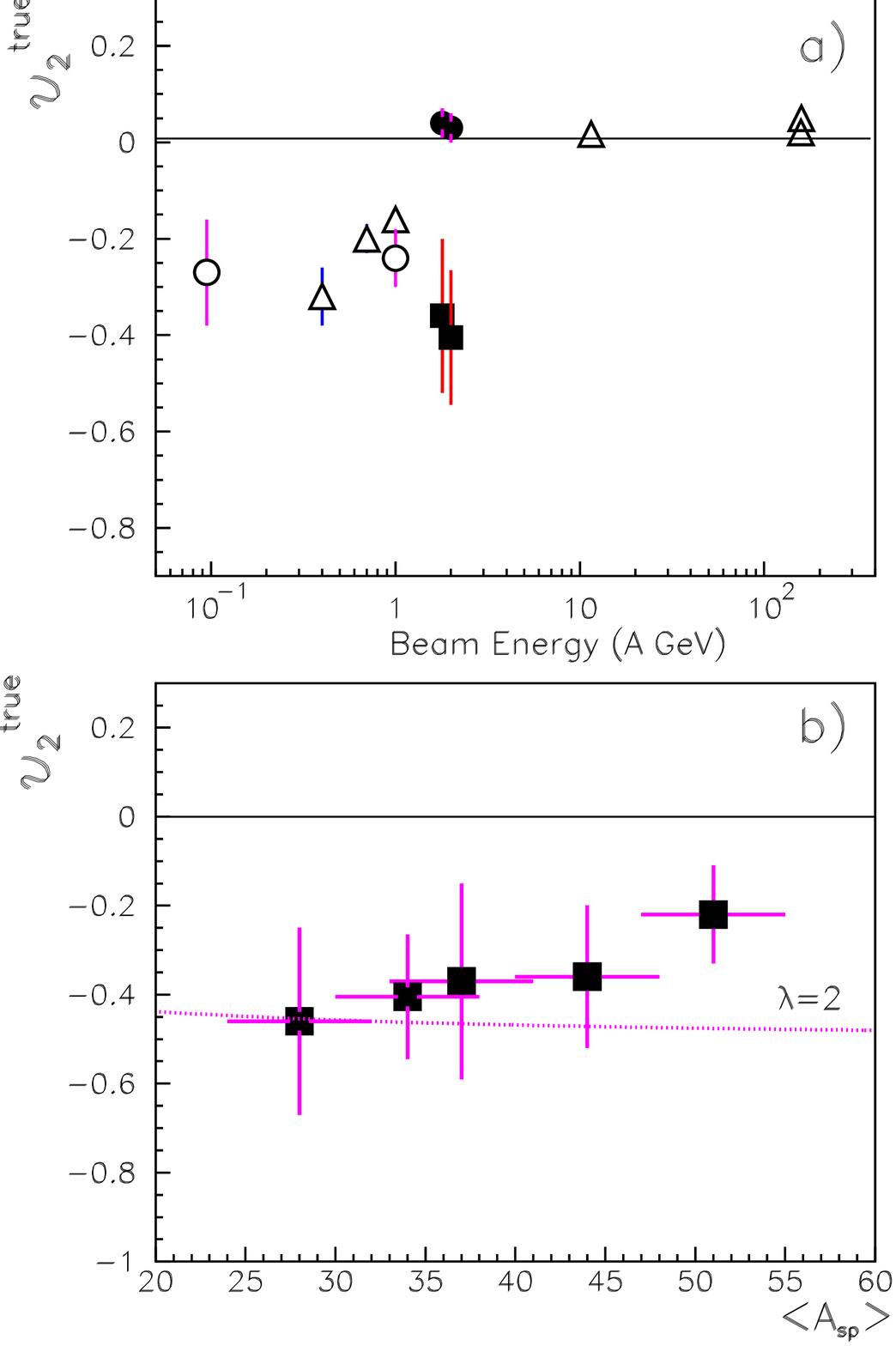}
      }
  \end{center}
  \vspace{-1.15cm}
  \begin{center}\parbox{14.0cm}{
     \caption{(a)Beam energy dependence of 
the elliptic flow signal $v_2^{true}$ for neutral (circles) 
and charged (triangles) pions and
$\eta$ mesons (squares) near midrapidity.
Open symbols correspond to data deduced from published anisotropies and full
symbols to our data (see text); 
(b)The parameter $v_{2}^{true}$ for $\eta$ mesons as a function
of the number of spectators $\langle A_{sp} \rangle$. 
The dash-dotted line represent results of the pure
geometrical absorption model (see text).}
     \label{fg:v2true}
  }\end{center}
  \vspace{-0.45cm}
\end{figure}

In order to compare the present results with previous measurements of elliptic flow
\cite{schu94,lars,brill97,bar99,agg99,app98} (see Table~\ref{tb:v2tru}), we plotted
the $v_2^{true}$ values for neutral and charged pions near
midrapidity as a function of the beam energy, see 
Fig.~\ref{fg:v2true}a. It can be seen that 
the pion elliptic flow 
undergoes a transition from out-of-plane to in-plane
emission around beam energy of 2 AGeV, 
which can be compared with theoretical prediction \cite{bao99}.
This effect can qualitatively be understood by taking into account the collision dynamics. 
The spectator nucleons leave the interaction region after the so-called passage time t$_{pass}$ of 
the order 2R/$\gamma\beta$ (where R is the nuclear radius and $\gamma$ is the Lorentz-contraction factor),
and the interaction with the spectator matter of particles produced later are not
significant \cite{olit,liu99,bao99,dan98}. Recent RQMD model calculations for the
Au+Au system at a beam energy of 2 AGeV \cite{liu99} 
show that the time-averaged values of $v_2$ for both pions are positive (indicating a preferred 
emission in the reaction plane), but pions emitted at the time scale less than 
t$_{pass}\simeq$13.5 fm/c  (see Table~\ref{tb:tpass}) exhibit negative elliptic flow. It is known that the 
freeze-out time of pions is strongly dependent on their transverse momentum and 
that most of the high p$_t$ pions freeze out early in the reaction \cite{bass98,sen99,bao99}.
Therefore, the observed nearly isotropic emission of neutral pions, except the high p$_t$ pions in 
light colliding systems at 2 AGeV, can qualitatively be understood.

\begin{table}
\begin{center}
 \begin{tabular}{|c|c|c|c|c|c|c|c|c|}\hline \hline
A+A & {\small E (AGeV)}& $\pi$ & centrality  & p$_t$ {\small (MeV/c)} & $v_{2}^{meas}$ & $\langle
cos2\Delta\phi\rangle$  & $v_{2}^{true}$  &  Ref.       \\ \hline \hline
Ar+Au & 0.095 &  $\pi^{0}$ & semiperiph.  & $\le$ 200 & -0.12$\pm$0.05& 0.44$^{1}$ & -0.27$\pm$0.11 &
 \cite{schu94}  \\ 
 &  &   & {\small Z$_{PLF}$=11-12}  & & &  &  &
   \\  
Ar+Au  &  0.095 &  $\pi^{0}$ & peripher.  & $\le$ 200 & -0.23$\pm$0.09& 0.44$^{1}$ & -0.52$\pm$0.21 &
 \cite{schu94}  \\ 
&   &  & {\small Z$_{PLF}$=14-17}   &  & &  &  &
  \\ \hline  
Ar+Al & 0.095 &  $\pi^{0}$ & semicentr..  & -- & -0.20$\pm$0.08& 0.49$^{1}$ & -0.42$\pm$0.16 &
 \cite{bad97}  \\ \hline
Bi+Bi & 0.4 &  $\pi^{-}$  & MUL2 & 360-600   & -0.15$\pm$0.02 & 0.35 $^{2}$ & -0.43$\pm$0.06 & 
 \\ 
 & &    & MUL3 & 360-600   & -0.13$\pm$0.02 & 0.41 $^{2}$ & -0.32$\pm$0.05 & 
\cite{brill97} \\ \hline
Bi+Bi & 0.7 &  $\pi^{-}$  & MUL2 & 360-600   & -0.12$\pm$0.01 & 0.49 $^{2}$ & -0.24$\pm$0.02 & 
\cite{brill97} \\ 
 & &    & MUL3 & 360-600   & -0.10$\pm$0.01 & 0.51 $^{2}$ & -0.20$\pm$0.02 & 
\\ \hline
Bi+Bi & 1.0 &  $\pi^{-}$  & MUL2 & 360-600   & -0.08$\pm$0.01 & 0.50 $^{2}$& -0.16$\pm$0.02 & 
 \\ 
 & &    & MUL3 & 360-600   & -0.09$\pm$0.01 & 0.55 $^{2}$ & -0.16$\pm$0.02 & 
\cite{brill97} \\ \hline
Au+Au & 1.0 &  $\pi^{0}$  & semicentr. & 400-600   & -0.19$\pm$0.07 & 0.71$^{1}$ & -0.27$\pm$0.10 & 
\cite{lars} \\ \hline
Au+Au & 11.5 &  $\pi^{-}$  & peripher. & 400-500   & -- & -- & 0.01-0.02 & 
\cite{bar99} \\ \hline
Pb+Pb & 158 &  $\pi^{\pm}$  & semicentr. & 400-600   & -- & -- & 0.02$\pm$0.01 & 
\cite{app98} \\ \hline
Pb+Pb & 158 &  $\pi^{+}$  & semicentr. & 50-800   & -- & -- & 0.05$\pm$0.02 & 
\cite{agg99} \\ \hline
 \end{tabular}
 \end{center}
 \caption{ Elliptic flow of pions for different beam energies. Comments: 
$^{1}$ - correction factors $\langle
cos2\Delta\phi\rangle$ were estimated from the simulations using the  published uncertainty
of the reaction plane determination. $^{2}$ -  correction factors taken from ref. \cite{brill96}}
 \label{tb:v2tru}
\end{table}

On the other hand, the observed strong negative elliptic flow of $\eta$ mesons indicates
that a significant part of them freezes out while the spectators are still close to the participant zone. 
The  magnitude of elliptic flow is comparable with that for pions observed in heavy colliding systems
at energies 0.1-.4 A GeV 
%\cite{lars,brill,brill97}, 
\cite{lars} -- \cite{brill97}, 
see Fig.~\ref{fg:v2true}a. However, the p$_t$ dependence is different, 
see Fig.~\ref{fg:v2tru_ept}. Within large error bars, the 
$v_{2}^{true}$ values for $\eta$ mesons do not vary with transverse momentum in contrast to the
ones of pions 
%\cite{lars,brill,brill97}. 
\cite{lars} -- \cite{brill97}. 
In order to estimate the effect due to "shadowing" by spectators 
one can apply the formula 
\begin{equation}
v_2^{true}=0.5(1-R)/(1+R)~~,~with~~~~ R=exp(L/\lambda),
\end{equation}
where $\lambda$  is the mean free path for $\eta$ mesons in cold spectator matter
($\lambda_{\eta}\approx$ 1-2 fm \cite{mami96})  and 
$L=2\cdot A_{sp}^{1/3}$~~fm is the mean thickness of spectator matter.
The resulting dotted line in Fig.~\ref{fg:v2true}b describes the data for  $\eta$ mesons fairly well.

\begin{figure}
  \vspace{0.0cm}
  \begin{center}
    \mbox{
     \epsfxsize=13.0cm
     \epsffile{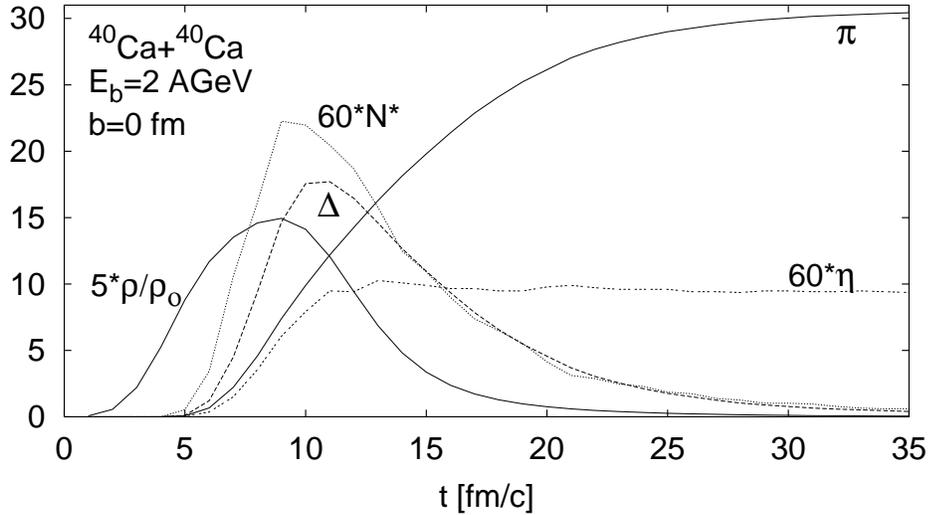}
      }
  \end{center}
  \vspace{-0.65cm}
  \begin{center}\parbox{14.4cm}{
     \caption{The prediction of BUU model for the time evolution of various quantities in a central collision of 
$^{40}$Ca+$^{40}$Ca at 2 AGeV \cite{wolfp}. 
}
     \label{fg:buu}
  }\end{center}
  \vspace{-0.45cm}
\end{figure}

The approximation by a shadowing scenario through cold nuclear matter for describing the strong
negative elliptic flow of $\eta$ mesons is also supported by different microscopic
model calculations, which show that $\eta$ mesons freeze out at earlier times 
as compared with pions \cite{gwolf,bao99,gudima}. 
As an example,  Fig.~\ref{fg:buu}  shows the prediction of the BUU model \cite{wolfp} for the time evolution of 
the $\Delta$, $N^*(1535)$, $\eta$ and $\pi$
yields in comparison to the density of the central region for central $^{40}$Ca+$^{40}$Ca collisions
at 2 AGeV. This figure shows that the $\eta$ mesons decouple from baryons much earlier than the pions.
Therefore, the shorter freeze-out time  may explain the much stronger azimuthal
anisotropy observed for $\eta$ mesons in comparison with pions. 

In summary, we have studied simultaneously the azimuthal angular distributions of $\pi^{0}$ 
and $\eta$ mesons emitted at midrapidity in the two colliding 
systems $^{58}$Ni+$^{58}$Ni at 1.9 AGeV and $^{40}$Ca+$^{nat}$Ca at 2 AGeV.
We observed a strong negative elliptic flow of  $\eta$ mesons. 
The elliptic flow of $\pi^{0}$ mesons is very weak in contrast to  data obtained for heavy 
colliding systems at 1 AGeV. The data can qualitatively be explained by the final state 
interaction of mesons in the spectator matter taking into account the collision dynamics.

This work was supported in part by the Granting Agency of the Czech Republic,
by the Dutch Stichting FOM, the French IN2P3, the German BMBF, 
the Spanish CICYT, by GSI, and the
European Union HCM-network contract ERBCHRXCT94066.

\begin{table}
\begin{center}
 \begin{tabular}{|c|c|c|c|c|c|}\hline \hline
System & R (fm)  & Energy (A GeV)& $~~\beta$(c)~~ &~~~$\gamma$~~~ & t$_{pass} (fm/c)$  \\ \hline
Ni+Ni  &  4.48       & 2.0  & 0.72 & 1.44 & 8.7 \\ \hline
Ca+Ca  &  3.96       & 2.0  & 0.72 & 1.44 & 7.7 \\ \hline
Au+Au  &  6.74       & 2.0  & 0.72 & 1.44 & 13.5  \\ \hline
Au+Au  &  6.74       & 1.0  & 0.59 & 1.24 & 18.5  \\ \hline
 \end{tabular}
\end{center}
\caption{ The nuclear radius ( R=1.6$\cdot$A$^{1/3}$ fm ), 
velocity $\beta$ (in units of the speed of light), Lorencz-contraction factor $\gamma$ and passage
time t$_{pass}$ for the systems Au+Au at 1-2 A GeV and Ni+Ni, Ca+Ca at 2 A GeV. All quantities are evaluated
in the nucleon-nucleon center-of-mass system.}
\label{tb:tpass}
\end{table}

\newpage

\newpage

\newpage

\newpage

\section{ Appendix {\bf A}: Flow of charged baryons}

\begin{figure}
  \vspace{-1.0cm}
  \begin{center}
    \mbox{
     \epsfxsize=13.0cm
     \epsffile{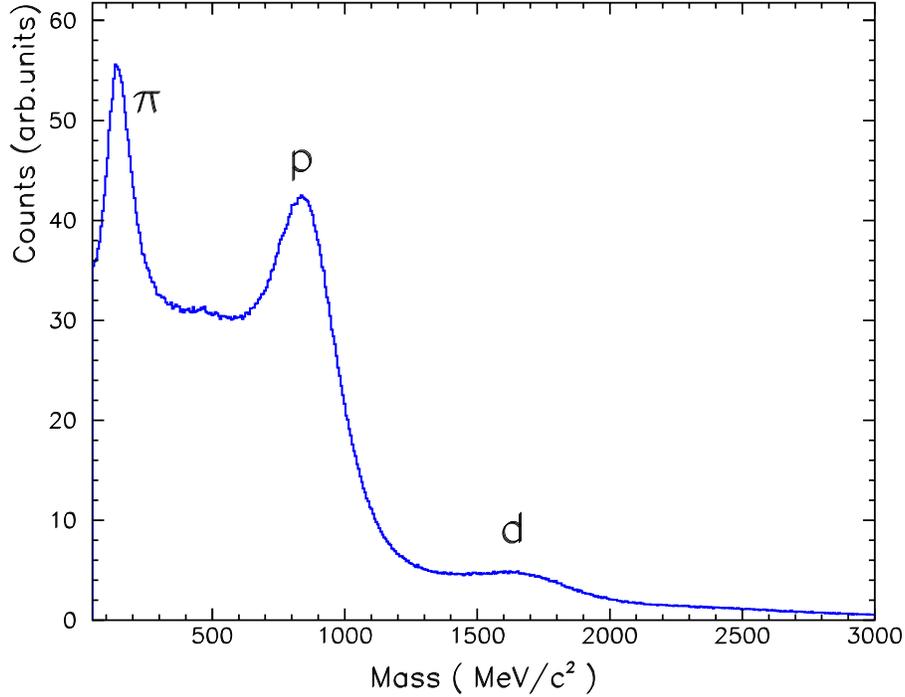}
      }
  \end{center}
  \vspace{-1.0cm}
  \begin{center}\parbox{13.7cm}{
     \caption{ Mass spectrum of charged particles detected by TAPS for the $^{58}$Ni+$^{58}$Ni system. 
Clearly separated peaks corresponding to charged pions, protons and deuterons are seen.}
     \label{fg:baryons}
  }\end{center}
  \vspace{0.0cm}
\end{figure}

\begin{figure}
  \vspace{-1.7cm}
  \begin{center}
    \mbox{
     \epsfxsize=13.4cm
     \epsffile{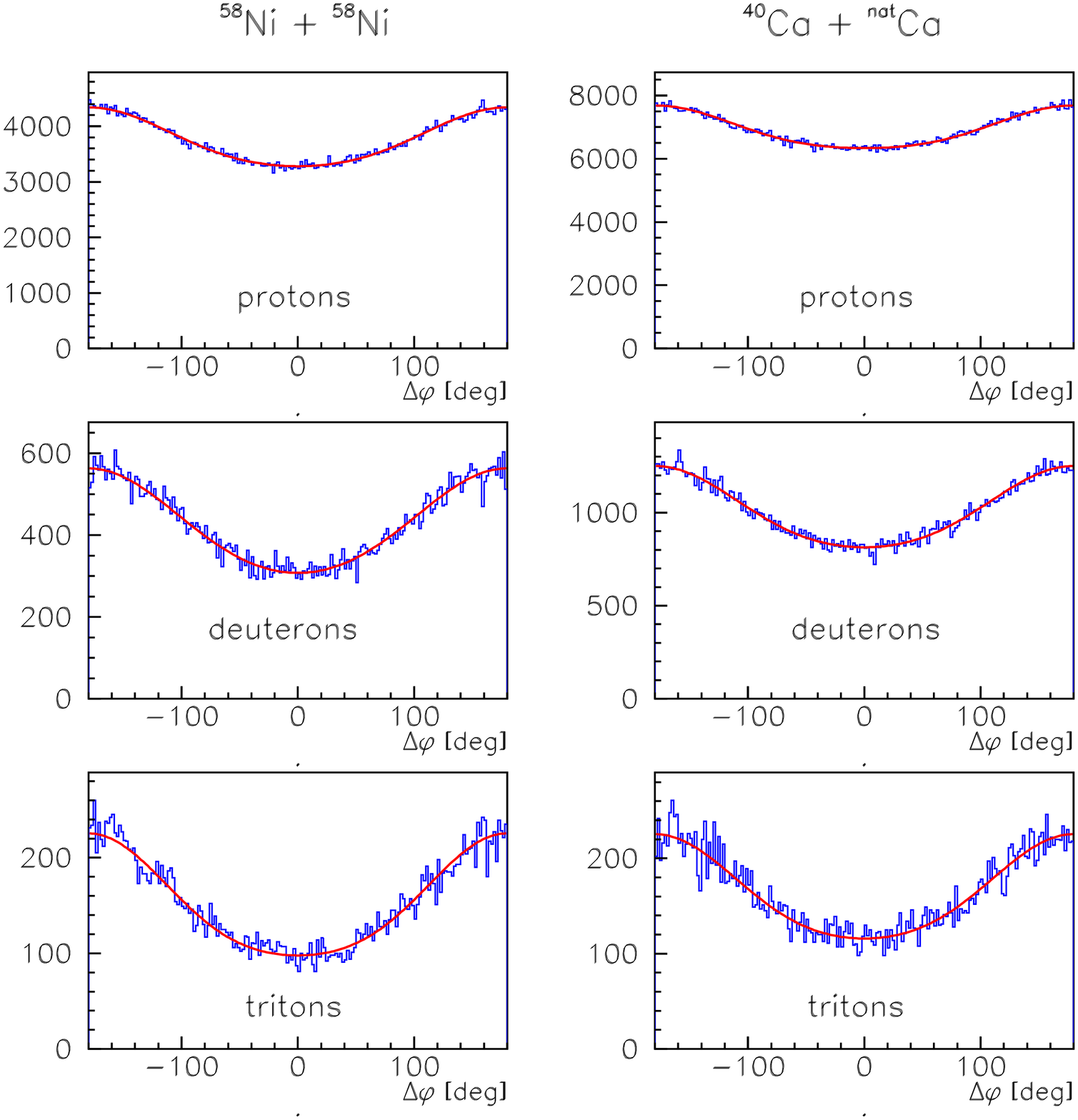}
      }
  \end{center}
  \vspace{-1.0cm}
  \begin{center}\parbox{14.5cm}{
     \caption{Azimuthal-angle distribution of charged baryons 
(protons, deuterons, tritons)
 at target-like rapidity (0.3 $<y_{lab}< 0.5 $). The left part contains 
the $^{58}$Ni+$^{58}$Ni data at 1.9 AGeV and the right part the $^{40}$Ca+$^{nat}$Ca data at 2 AGeV.}
     \label{fg:bar_flow}
  }\end{center}
  \vspace{0.0cm}
\end{figure}

As TAPS allows the identification of charged particles by time-of-flight 
discrimination, pulse-shape discrimination and the charged-particle veto 
counters, we used this feature to verify our method of the reaction-plane reconstruction.
As an example, Fig.~\ref{fg:baryons} shows the mass spectrum of charged particles detected
by TAPS for the reaction $^{58}$Ni+$^{58}$Ni at 1.9 AGeV.
 In this figure clearly separated peaks corresponding to charged pions, protons and
deuterons are seen. 
Fig.~\ref{fg:bar_flow} shows the experimental azimuthal-angle distributions of identified charged baryons 
(protons, deuterons and tritons) detected in TAPS with respect to the reaction 
plane for both studied reactions. 
The directed flow of charged baryons in the target-like rapidity region 
(the preferential emission in the reaction plane) is clearly observed.  
This verification demonstrates the quality of the reaction-plane determination for both colliding systems. 

\end{document}